\documentclass[reprint, amsmath,amssymb,aps,prx,nofootinbib]{revtex4-2}

\usepackage{amsmath}

\usepackage{amsfonts}
\usepackage{hyperref}

\usepackage{graphicx}
\usepackage{hyperref}

\usepackage{url}

\usepackage{booktabs}
\usepackage{tabularx}
\usepackage{array}

\begin{document}

\title{Temperature Beyond Equilibrium\\
in Isolated Quantum Many-Body Systems and Their Subsystems}

\author{Maurizio Fagotti$^{1}$}

\affiliation{Universit\'e Paris-Saclay, CNRS, LPTMS, 91405, Orsay, France}

\email{maurizio.fagotti@universite-paris-saclay.fr}

\begin{abstract}
Temperature is one of the central concepts of thermodynamics, yet its meaning far from equilibrium remains unclear. The problem is especially challenging in isolated quantum many-body systems, whose states evolve unitarily, may remain far from equilibrium, and retain energy coherence, a genuinely quantum feature with no direct classical counterpart. Specifically, energy fluctuations in a nonstationary quantum state have two distinct origins. Part of these fluctuations comes from uncertainty in the energy populations and has the usual thermodynamic meaning. The rest comes from quantum coherence between energy sectors and is responsible for the state's time dependence. We propose that, even away from equilibrium, temperature identifies the state within the family of regular states sharing the same energy-coherence structure. The resulting inverse temperature is obtained from a principle of minimum discrimination information, which selects the least distinguishable state compatible with the prescribed energy and coherence structure. This replaces the usual maximum-entropy principle, which, applied naively, can yield a biased estimator of the inverse temperature. We also lay the groundwork for an extension of the construction to subsystems and show that, although the inverse temperature of a subsystem is not determined by its reduced state alone, under an alignment assumption, its instantaneous rate of change is a local quantity for sufficiently large subsystems. 
\end{abstract}
\maketitle

\section{Introduction}

Nearly two centuries after the first explicit proposal of an absolute thermodynamic temperature scale, there is still no universally accepted notion of nonequilibrium temperature~\cite{CasasVazquezJou2003,Puglisi2017Temperature,Zhang2019Local}. Once local equilibrium breaks down, the reasoning that makes temperature unique in equilibrium no longer applies, and different extensions of the concept need not agree. Nonequilibrium temperatures have therefore been defined through thermal contact with equilibrium reservoirs~\cite{Muschik1977A}, fluctuation-dissipation relations~\cite{Cugliandolo1997Energy,Kurchan2005In,Cugliandolo2011The}, weak probe thermometers~\cite{Meair2014Local,Ye2016Thermodynamic}, entropy maximization with selected relevant variables~\cite{Alipour2021Temperature}, kinetic averages~\cite{Goldhirsch2008Introduction}, local thermal observables~\cite{Buchholz2002Thermodynamic}, or late-time local equivalence with equilibrium states~\cite{Rigol2008Thermalization,DAlessio2016From}. These constructions extend different roles played by temperature at equilibrium: contact-based definitions depend on which exchanges are suppressed, fluctuation-dissipation temperatures can depend on the observable and timescale considered, and entropy-maximization approaches depend on the constraints chosen to describe the state. Away from equilibrium, there is therefore no general reason for the resulting temperatures to coincide. Yet temperature remains one of the basic quantities through which physical states are characterized, including outside strict equilibrium.

This ambiguity may stem from treating nonequilibrium temperature as a remnant of equilibrium rather than as an intrinsic property of a state together with its dynamics. A more general framework should instead make equilibrium thermodynamics emerge as a special case. This requires more than assigning a scalar parameter to an arbitrary state. Equilibrium temperature is meaningful because thermodynamics discards microscopic information: once microscopically distinct states have been identified with one another, it is the coordinate that locates a state within the family that remains. The central question is therefore which structure can play this organizing role away from equilibrium.

At equilibrium, stationarity provides the basic constraint~\cite{HaagHugenholtzWinnink1967}. Within the stationary sector, microscopic complexity can collapse into a thermodynamic description. A paradigmatic manifestation is the eigenstate thermalization hypothesis~\cite{Deutsch1991Quantum,Srednicki1994Chaos,Rigol2008Thermalization,DAlessio2016From}, according to which local observables in individual finite-energy-density eigenstates approach their equilibrium values in the thermodynamic limit.  Nonstationary states are usually regarded as both physically richer and intrinsically less amenable to an analogous thermodynamic organization. Here we question whether the distinction between stationary and nonstationary states should be regarded as a fundamental dichotomy.

Our starting point is to reinterpret stationarity as the absence of coherent energy fluctuations. Coherent energy fluctuations constitute the part of the energy uncertainty that cannot be understood as classical ignorance about energy populations~\cite{Streltsov2017Coherence}: they quantify coherence between different energy eigenspaces, precisely the component that allows the state to evolve under the Hamiltonian~\cite{Lostaglio2015TimeTranslation,Marvian2016Speed}. While local thermalization is often accompanied by local decoherence~\cite{Polkovnikov2011Colloquium,DAlessio2016From}, the coherent energy content of the full state of an isolated system is conserved by unitary dynamics. Once this information is retained, the sharp distinction between stationary and nonstationary states gives way to a richer classification. This is the perspective of Ref.~\cite{Fagotti2026Quantum}, which introduced a minimum-variance foliation grouping states by their energy-coherence structure and proposed a leaf-typicality hypothesis extending the logic of eigenstate thermalization beyond the stationary sector. 

Here we show that this organization supports an intrinsic thermodynamics of isolated quantum spin chains. In the thermodynamic limit, regular states sharing the same energy-coherence structure are generically connected by a distinguished thermodynamic curve analogous to the family of Gibbs states, and inverse temperature emerges as the coordinate along it.  Away from equilibrium, however, quantities that coincide at equilibrium can become distinct: e.g., inverse temperature can differ from the thermodynamic-entropy response to an energy change, and the inverse temperature assigned to a sufficiently large subsystem cannot in general be inferred from the reduced state alone, while its instantaneous rate of change can nevertheless become a local quantity.

The global construction is established in a controlled thermodynamic-limit regime. Its extension to subsystems, on the other hand, involves an additional challenge. We show that the thermodynamic structure of the full system induces a corresponding structure on its parts, with boundary effects accounting for the ambiguity associated with the finite size of a subsystem. To obtain a compact dynamical equation for the subsystem inverse temperature, however, we rely on an additional assumption concerning the relation between the physical evolution and the induced thermodynamic structure, which extends beyond the organizational scheme developed here.

\subsection{Physical setting}~\label{ss:initialstate}

We aim for a presentation that remains broadly accessible.  Therefore, we draw analogies with the finite-volume formalism as long as this does not obscure the thermodynamic limit. In addition, the analysis will be restricted to quantum spin chains, where short-range systems enjoy strong regularity properties~\cite{Araki1969Gibbs,BluhmCapelPerezHernandez2022}. We nevertheless formulate the construction in a way that should facilitate a future fully $C^\star$-algebraic treatment.

Our starting point is a quantum spin chain whose finite-volume time evolution, on an interval $I\subset\mathbb Z$ of length $|I|$, is generated by a Hamiltonian
\begin{equation}
H_0^{(I)}=\sum_{\ell\in I}h_{0,\ell},
\end{equation}
where $h_{0,\ell}$ is localized around site $\ell$ and the interaction decays sufficiently fast. The corresponding dynamics is
\begin{equation}
\sigma_0^t(O)=e^{iH_0^{(I)}t}Oe^{-iH_0^{(I)}t}.
\end{equation}
In the thermodynamic limit $|I|\to\infty$, or $I\nearrow\mathbb Z$, we use the same notation $\sigma_0^t$ for the induced time-translation automorphism of the quasilocal algebra.

We denote the expectation value of a local observable $O$ in the state $\omega$ either by $\mathrm{tr}(\rho_\omega O)$ or by $\omega(O)$, depending on whether we want to emphasize the finite-volume density-matrix picture or the functional representation of the state.

The initial state is assumed to be an equilibrium state for the dynamics $\sigma_0^t$. In finite volume, it is represented by a Gibbs ensemble
\begin{equation}
\rho_\omega\propto e^{-\beta_0 H_0^{(I)}}\, .
\end{equation}
In the thermodynamic limit, we assume that it satisfies the Kubo--Martin--Schwinger (KMS) condition \cite{BratteliRobinson1997}: for every pair of local observables $A,B$, there exists a function $F_{A,B}(z)$, holomorphic in the strip $0<\mathrm{Im}z<\beta_0$ and continuous on its closure, such that
\begin{equation}
\begin{aligned}
F_{A,B}(t)&=\omega\bigl(A\sigma_0^t(B)\bigr),\\
F_{A,B}(t+i\beta_0)&=\omega\bigl(\sigma_0^t(B)A\bigr).
\end{aligned}
\end{equation}
For one-dimensional quantum spin chains with sufficiently local interactions, the KMS condition characterizes the equilibrium state at fixed inverse temperature \cite{Araki1969Gibbs,Araki1974Equivalence,Araki1975Uniqueness, Kishimoto1976Uniqueness}. In this setting, KMS states also satisfy a local equilibrium condition,  according to which finite subsystems take Gibbs form once the boundary coupling is properly taken into account \cite{Araki1974Equivalence,Araki1975Uniqueness,Kishimoto1976Uniqueness}. Thus, the KMS condition gives meaning to the state $\omega$ directly in the thermodynamic limit and characterizes it through the pair $(\beta_0,\sigma_0^t)$.

The basic physical setting of our investigation is a quantum quench, with a local post-quench dynamics $\sigma^t$, which in finite volume is given by
\begin{equation}\label{eq:sigmat}
\sigma^t(O)=e^{iH^{(I)}t}Oe^{-iH^{(I)}t}\, ,\quad \text{with}\quad
H^{(I)}=\sum_{\ell\in I}h_\ell ,
\end{equation}
where $h_\ell$ is localized around site $\ell$. Since the dynamics is quasilocal, the time-evolved state remains well defined in the thermodynamic limit at every finite time.

\section{Summary of the results}

For quick reference and guidance, we summarize here the main results of the paper, presenting first the physical meaning and then highlighting the underlying structure.  For the purposes of this section, a ``leaf'' may be taken informally to be the family of states sharing the same energy-coherence structure; in particular, the states commuting with the Hamiltonian form a leaf, which we call the commuting leaf. The terminology is made precise in Section~\ref{s:mVfoliation}. 

\subsection{Nonequilibrium temperature}
\emph{We develop a notion of temperature for isolated quantum spin chains out of equilibrium. In equilibrium, changing temperature moves the state along the thermal curve of Gibbs states. We show that an analogous curve emerges also out of equilibrium, once the coherent energy content of the state is kept fixed. For states with well-behaved local properties, such as clustering correlations, the position on this curve is interpreted as the inverse temperature.}

We establish the nonequilibrium analogue of equilibrium thermal states, building on the flow perspective of Ref.~\cite{Doyon2017Thermalization}. At equilibrium, that is, on the commuting leaf, thermal states $\omega_\beta$ may be viewed as the states reached from the infinite-temperature state by the energy flow ($(\omega_{\beta_0},\Delta \beta)\mapsto \omega_{\beta_0+\Delta \beta}$). This flow is well defined in the thermodynamic limit and, at sufficiently high temperature, preserves exponential clustering.

We show that an analogous construction exists on non-commuting leaves---Section~\ref{s:isolated}. The inverse temperature becomes the coordinate of a canonical ``pseudolocal'' flow on the leaf, whose infinitesimal action on local observables is represented by (at equilibrium $\tilde H_{[\omega]}=0$) 
\begin{equation}\label{eq:infaction}
\widehat H_{\omega}(O)=\langle\!\langle H,O\rangle\!\rangle_{\omega}^{c}+\frac{i}{2}\omega\bigl([\tilde H_{[\omega]},O]\bigr)\, .
\end{equation}
The right-hand side is understood as the thermodynamic limit of the corresponding finite-volume actions. Here $\langle\!\langle \cdot,\cdot\rangle\!\rangle_\omega^c$  denotes the symmetrized connected correlation
\begin{equation}\label{eq:symconncorr}
\begin{aligned}
\langle\langle A,B\rangle\rangle^c_\omega=&\frac{1}{2}(\langle A,B\rangle^c_\omega+\langle B,A\rangle ^c_\omega)\\
\langle A,B\rangle^c_{\omega}=&\omega(AB)-\omega(A)\omega(B)\\
\end{aligned}
\end{equation}
and $\tilde H_{[\omega]}$ (defined in \eqref{eq:tildeHomega}) is one half of the symmetric logarithmic derivative associated with the infinitesimal adjoint action generated by $H$. Within the regularity regime established in Appendix~\ref{a:leafthrough}, the flow remains, in a neighborhood of the given state $\omega$, within the class of exponentially clustering states. This statement can be taken as the working, non-technical meaning of pseudolocality used here: the flow preserves the local regularity properties of the state. For a rigorous formulation in terms of pseudolocal charges and their Hilbert-space structure, we refer the reader to Ref.~\cite{Doyon2017Thermalization} (the leaf generalization of the inner product can be found in Section~\ref{ss:inner}).

We stress a shift of perspective, analogous to the one underlying statistical descriptions of late-time local observables after global quenches, where the ensemble is reconstructed from the integrals of motion rather than fixed a priori~\cite{EsslerFagotti2016,VidmarRigol2016}. At equilibrium, temperature is often specified first and used to define the canonical state. Here the order is reversed: the state is given, together with the dynamics, and the problem is to assign a temperature to it. The inverse temperature is identified with the coordinate along the canonical flow connecting the state to the appropriate reference point on its leaf, which plays the role of the tracial state on the commuting leaf (in finite volume, the latter is the state with density matrix proportional to the identity).

The orbit of the canonical flow is obtained using a principle of minimum discrimination information. Remarkably, Section~\ref{s:examples} provides numerical evidence that the latter  might not be equivalent to the principle of maximum entropy in which the states on the leaf are qualified only geometrically as points on the simplex. That is, the inverse temperature does not match the derivative of the thermodynamic entropy---Eq.~\eqref{eq:thentropy}---with respect to energy or, alternatively, the parameter $\beta$ of the leaf canonical ensemble of Ref.~\cite{Fagotti2026Quantum}, Eq.~\eqref{eq:leafGibbs}. 
We point out that the potential disagreement with the principle of maximum entropy can also be viewed constructively---Appendix~\ref{a:principle}. The entropy--energy relation can in fact be restored by measuring the thermodynamic entropy relative to the reference state at zero inverse temperature. This, however, requires a distinguished regular reference state on the leaf, which, to the best of our understanding, the leaf structure alone does not provide straightforwardly. 

\subsection{Generic and non-generic leaves}
\emph{In equilibrium, additional conservation laws, integrability, structural constraints, or phase transitions signal that a single smooth temperature coordinate may not be enough. We show that the same question can be asked on every leaf. Generic leaves support only the canonical temperature direction; non-generic leaves may support extra thermodynamic directions or decompose into distinct sectors.}

We detach the notion of genericity from the Hamiltonian, or from the dynamics it generates, and attach it instead to the leaves of the foliation. In this language, what is usually meant by a ``generic Hamiltonian'' or a ``generic dynamics'' corresponds to genericity of the commuting leaf.  A leaf of the minimum-variance foliation is generic if it supports a single pseudolocal thermodynamic direction, namely the canonical one---Section~\ref{s:generic}.

We expect the set of states lying on generic leaves to be dense in the regular state space. Nevertheless, some leaves can support additional pseudolocal flows---Section~\ref{ss:nongeneric}. Such non-generic leaves naturally give rise to leaf grand-canonical ensembles or, when infinitely many independent pseudolocal charges are present, to leaf generalized Gibbs ensembles.  In other cases, the obstruction may be geometric: the leaf may fragment into distinct thermodynamic sectors, or the canonical trajectory on the leaf may become nonanalytic, signalling a phase transition within the leaf. This can occur even when the physical dynamics generated by $H$ is, for all practical purposes, generic---see Section~\ref{s:examples} for examples.

\subsection{Subsystem temperature}
\emph{
We lay the groundwork for extending the notion of temperature to parts of the system. For a finite subsystem $A$, boundary effects produce an intrinsic ambiguity of order $1/|A|$ in its inverse temperature, so the latter becomes asymptotically well defined as the subsystem size increases. Out of equilibrium, there is a further complication: its value cannot, in general, be inferred from the state of the subsystem at a single time. Nevertheless, when the subsystem is large enough for a single temperature to provide a meaningful thermodynamic description, its instantaneous rate of change can become a local quantity, determined by the state and its evolution in the vicinity of the subsystem.
}

Despite subsystems being generally open systems, the global canonical pseudolocal flow induces an effective local thermodynamic direction on subsystems---Section~\ref{s:subsystems}.  This allows us to construct an effective local foliation that, under an ``alignment assumption'', determines  the time derivative of the subsystem inverse temperature. Along the physical orbit of the subsystem state, this can be written as
\begin{equation}\label{eq:firstlaw0}
\frac{d}{dt}\Bigl[ \omega\left(K_A^{\mathcal L,\omega}\otimes \mathbf 1_{\bar A}\right)\Bigr]+\widehat H_{\omega}\left(K_A^{\mathcal L,\omega}\otimes \mathbf 1_{\bar A}\right)\frac{d  \beta_A}{dt}=0\, .
\end{equation}
Here  $K^{\mathcal L,\omega}_A$ is the leaf subsystem modular Hamiltonian, which is defined in Section~\ref{ss:lowsub} and is a close cousin of the entanglement Hamiltonian.   The first term is the time variation of the thermodynamic entropy---Eq.~\eqref{eq:thentropy}---which is expected to be a boundary contribution. The factor multiplying $\frac{d \beta_A}{dt}$ plays the role of a heat capacity associated with the subsystem, which is instead extensive. Thus, it is reasonable to expect that the inverse temperature varies over macroscopic times, i.e.,  times that grow with the subsystem size.  

The time derivative of the subsystem inverse temperature can be integrated once a reference value has been fixed. We argue that the subsystem inverse temperature has an intrinsic indetermination $O(|A|^{-1})$, hence any time at which $\beta_A$ can be determined with that degree of accuracy can be taken as reference time. We show that this is possible when the canonical flow induces a flow on the subsystem that, roughly speaking, is canonical in the bulk, as clarified in Section~\ref{ss:equi}. In common situations, this happens at the initial time, since the state is usually constructed with strong local properties. Alternatively, in thermalizing situations, a natural reference is the asymptotic regime $t\to\infty$, where the reduced state of the subsystem becomes stationary.  The reference inverse temperature is then supplied by the coordinate of the corresponding weak-canonical-flow orbit.

A final comment concerns the alignment assumption used to derive \eqref{eq:firstlaw0}. While its range of validity requires further investigation, without this assumption the equation for $\frac{d\beta_A}{dt}$ would acquire additional terms and possibly depend on additional quantities. Equation \eqref{eq:firstlaw0} can therefore also be regarded as the minimal evolution equation for the subsystem inverse temperature.

\section{The minimum-variance foliation}\label{s:mVfoliation}

\begin{figure}
\centering
\includegraphics[width=0.45
\textwidth]{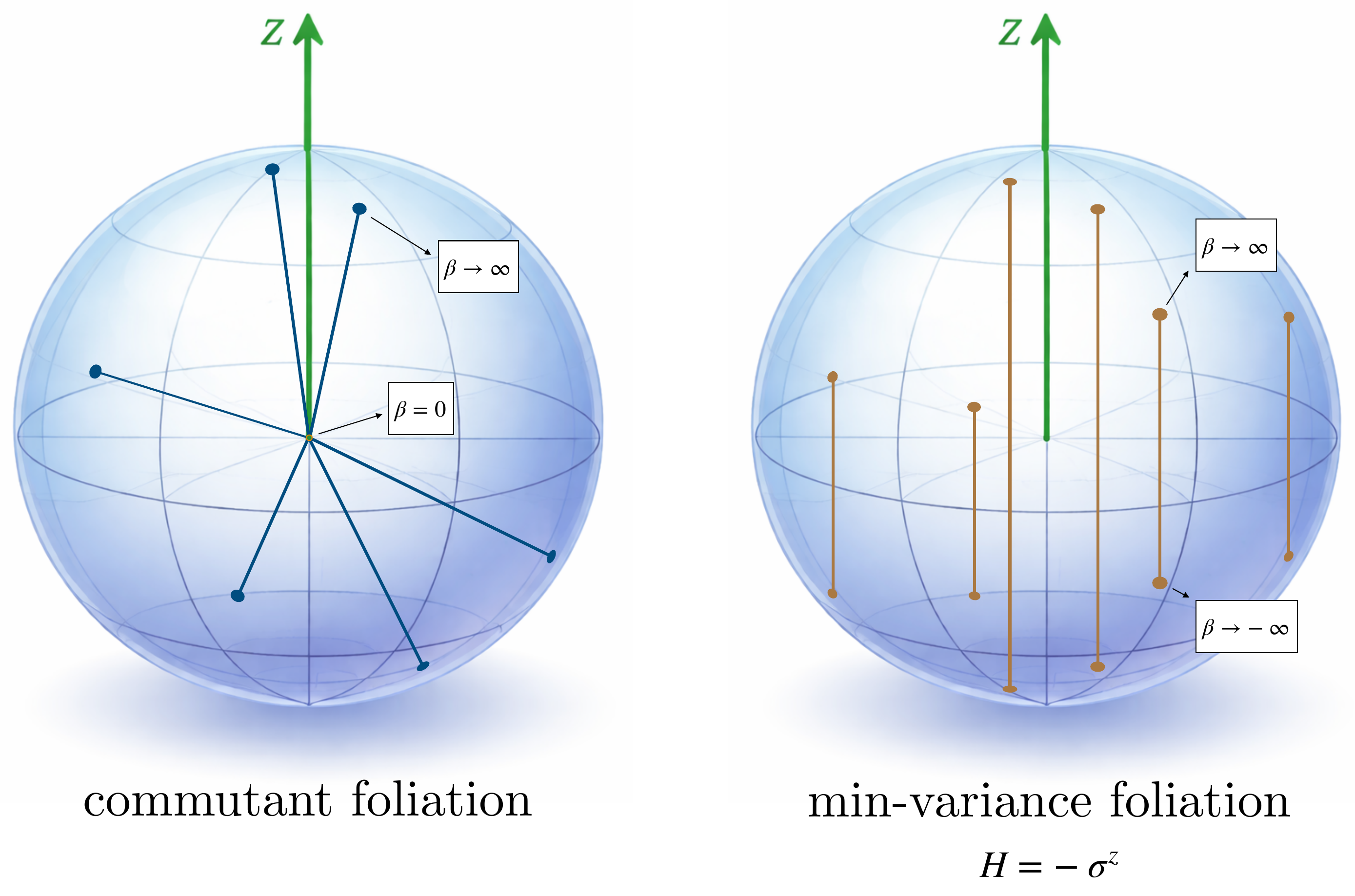}
\caption{
\textbf{Two foliations of state space.}
The cartoon compares the commutant foliation and the minimum-variance foliation of the state space of a two-dimensional Hilbert space. The state space is represented by the Bloch ball. The commutant foliation is defined on the state space with the tracial state $\frac{1}{2}\mathrm{I}$ removed; its leaves are half-open radial segments. By contrast, the leaves of the minimum-variance foliation associated with the Hamiltonian $H=-\sigma^z$ are closed chords parallel to the $z$ axis. The inverse temperature plays the role of coordinate on the leaf. 
}\label{f:foliation}
\end{figure}

To make the connection between the standard viewpoint and the framework proposed in this paper more explicit, we start with reinterpreting the equilibrium thermodynamics geometrically. 

We regard the state of the system as a point in state space. In a $d$-dimensional Hilbert space, this point can be specified by $d^2-1$ real parameters, for instance by expanding the density matrix in a basis of traceless Hermitian operators. Not every choice of these parameters, however, corresponds to a physical state: positivity imposes nonlinear constraints. In such situations, it is useful to organize the state space by separating the description of a state into two pieces of information: the \emph{leaf} to which the state belongs, and the position of the state within that leaf. The leaf identifies the class of states that share the structural property used to define the organization, while the position on the leaf specifies the remaining degrees of freedom. When these leaves fit together smoothly, covering the state space without overlaps, one obtains a foliation. Choosing a foliation is analogous to choosing a way of organizing the books in a library: many choices are possible, but some make the collection more accessible than others.

\begin{figure}
\centering
\includegraphics[width=0.23
\textwidth]{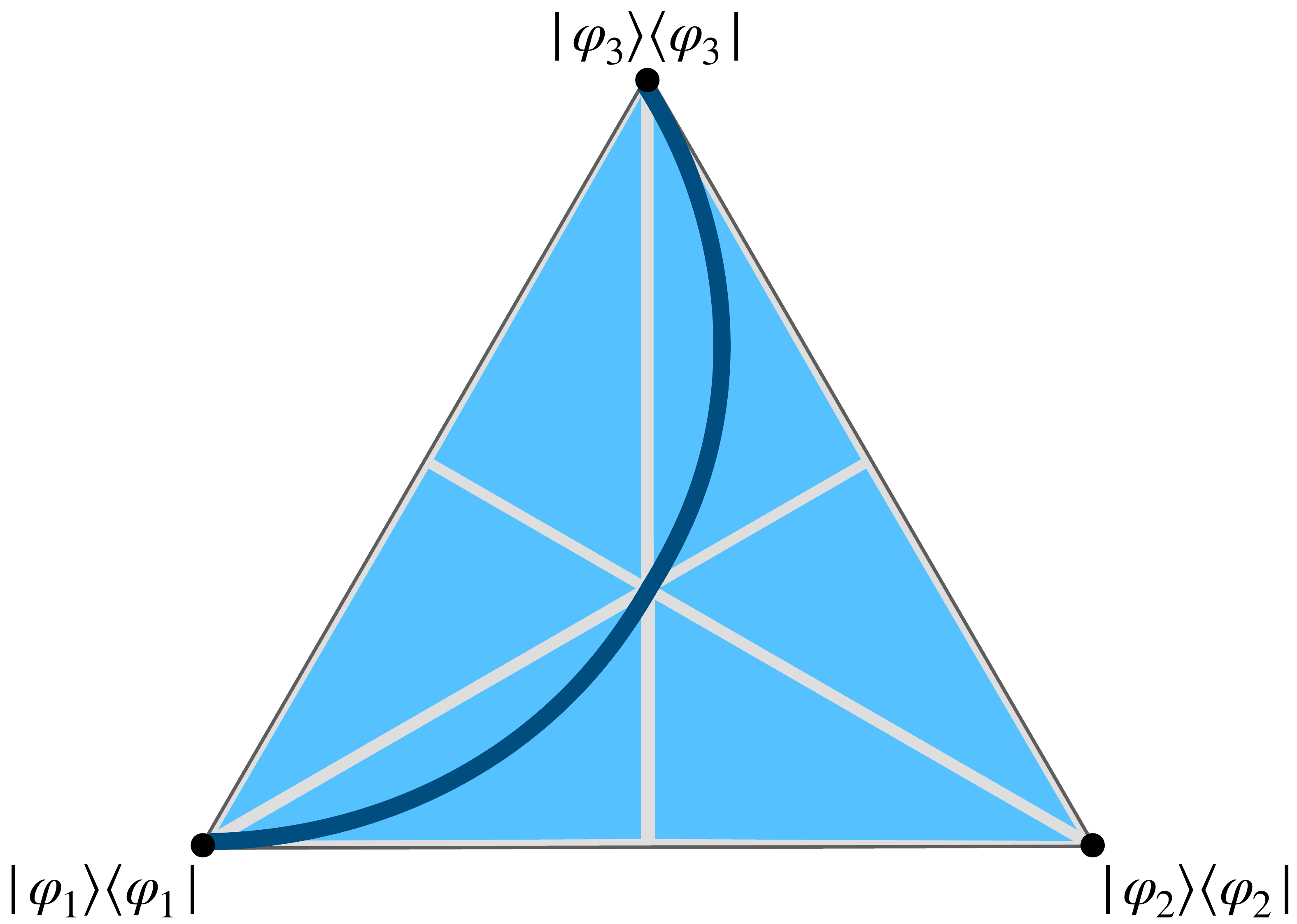}
\hspace{0.1cm}
\includegraphics[width=0.23
\textwidth]{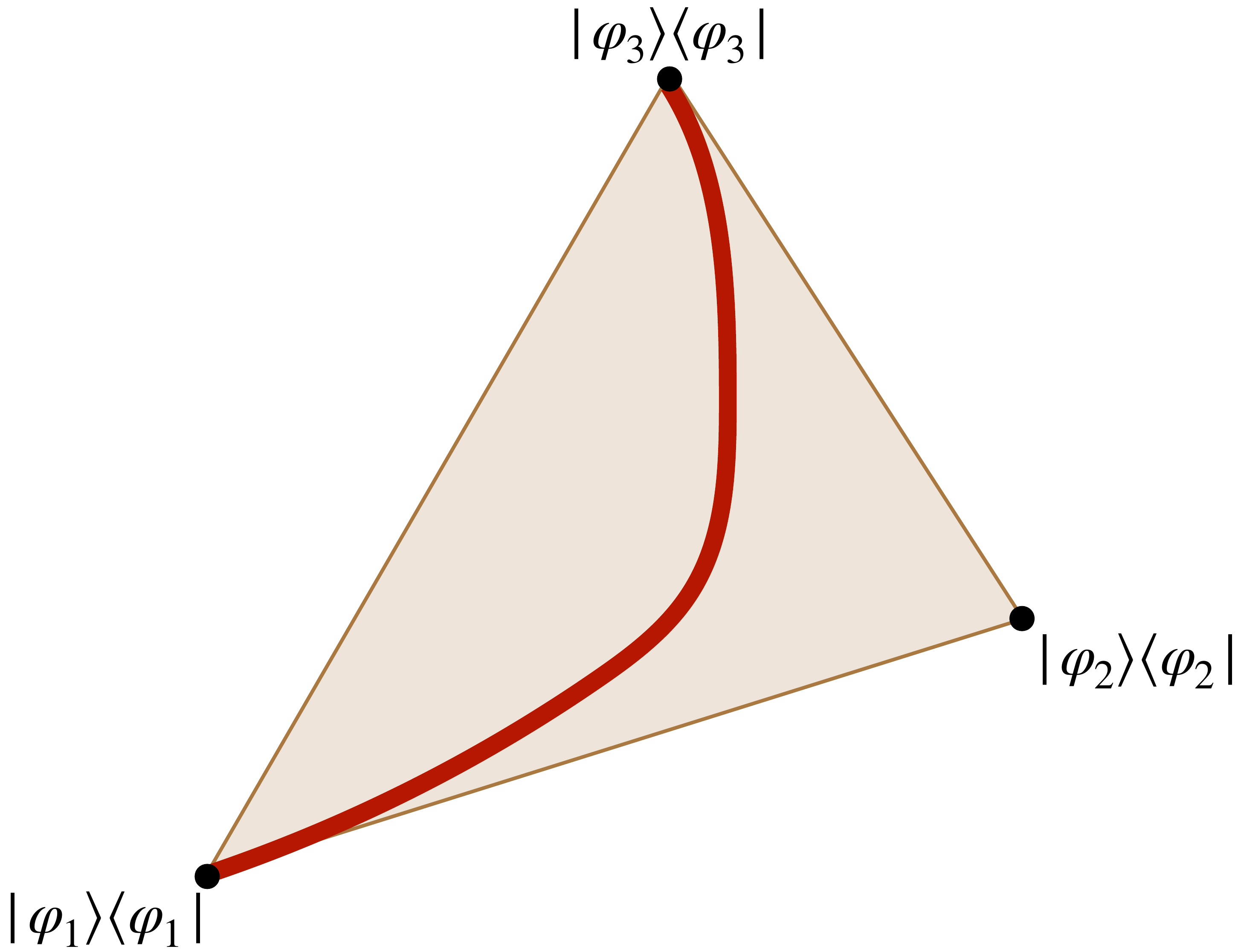}
\caption{
\textbf{From foliations to thermodynamics.}
\textsc{Left}: The six leaves of the commutant foliation containing stationary states of $H_3=\sum_{i=1}^3 \mathcal E_i \lvert \varphi_i\rangle\langle \varphi_i\rvert$, drawn as a cartoon of the many-body case. Each leaf is two-dimensional, but in a many-body system the eigenstate thermalization hypothesis reduces the states relevant for local observables to a one-dimensional family, represented here by the thick curve joining the ground state to the highest-energy one, along which the inverse temperature is the coordinate.
\textsc{Right}: The same on a leaf of the minimum-variance foliation, where the pure states of the decomposition are not  orthogonal. Under leaf typicality the reduction again selects a one-dimensional family; which curve on it carries the thermodynamic coordinate is the subject of Section~\ref{s:isolated}.
}\label{f:commutant}
\end{figure}

Returning to equilibrium states, the characterization through the pair $(\beta_0,\sigma_0^t)$ can be viewed as arising from a foliation of state space, which we call the \emph{commutant foliation}. This foliation is defined on the regular set of density matrices with simple spectrum. Its connected leaves consist of density matrices that commute with one another and share the same ordered eigenbasis. Thus the commutant foliation separates the data specifying the basis from the data specifying the populations in that basis---see Figure~\ref{f:foliation} left. Within this picture, the dynamics $\sigma_0^t$, through its generator $H_0^{(I)}$, selects the relevant leaf: equilibrium states commute with $H_0^{(I)}$ and are therefore diagonal in its eigenbasis. The Gibbs family then selects a one-dimensional curve on this leaf. Once this picture is supplemented by the eigenstate thermalization hypothesis, the inverse temperature $\beta_0$ is identified as the thermodynamically relevant coordinate along that curve---see Figure~\ref{f:commutant} left.

\subsection{Finite-volume construction}\label{ss:min-var}

After the quench, the pre-quench dynamics is decoupled from the actual time evolution, hence the scale of the pre-quench interaction becomes conventional: a rescaling of $\beta_0$ can be compensated by a rescaling of $H_0$. Thus $\beta_0$ no longer has an intrinsic meaning as the temperature of the state\footnote{This might remind the reader, who is familiar with the algebraic formalism, of the Tomita--Takesaki theory: the pair $(\beta_0,\sigma_0^t)$ is replaced by the modular automorphism group of the state, so that the inverse temperature is absorbed into the definition of the modular dynamics \cite{Takesaki1970,BratteliRobinson1997} (which, roughly speaking, is generated by the modular---or entanglement---Hamiltonian $-\log\rho_\omega$).}. We remark that such ambiguity was used in Ref.~\cite{ConnesRovelli1994} to link the notion of time to the state.  In the nonequilibrium setting considered here, the ambiguity is read in the opposite direction: once the system is put out of equilibrium, the leaf coordinate of the commutant foliation loses the direct thermodynamic meaning that it had at equilibrium. A different foliation, adapted to the actual nonequilibrium dynamics, is therefore needed in order to define a nonequilibrium notion of temperature. This is where the minimum-variance foliation proposed in Ref.~\cite{Fagotti2026Quantum} stands out.

The minimum-variance foliation separates the part of the energy uncertainty that can be attributed to classical population fluctuations from the residual coherent contribution. The physical intuition is that a nonstationary state cannot be written as a convex combination of energy eigenstates, but it can still be written as a convex combination of pure states chosen so as to minimize the decomposition-averaged energy variance. More precisely, a set of pure states $\{\varphi_i\}_{i=1}^d$ defines a candidate leaf if every state $\sum_i p_i |\varphi_i\rangle\langle\varphi_i|$ on that leaf admits no alternative pure-state decomposition with a smaller average variance of $H^{(I)}$. The coefficients $p_i$ are then the leaf coordinates, generalizing the energy populations of the commuting leaf. To the best of our knowledge, this optimal decomposition first appeared in Ref.~\cite{Yu2013Quantum}, where it was used to prove the conjecture of Ref.~\cite{Toth2013Extremal} relating the quantum Fisher information (QFI) to the convex roof of the variance. Readers unfamiliar with pure-state ensemble decompositions may think of them as different ways of preparing the same density matrix. If the preparation record is ignored, all decompositions leading to the same density matrix are equivalent for expectation values. If the record is retained, however, the decomposition carries additional classical information about the preparation. Convex-roof extensions exploit precisely this structure by optimizing averages over decompositions, and are widely used to construct entanglement measures in quantum information theory~\cite{Bennett1996Mixed,Vidal2000Entanglement,PlenioVirmani2007}.

For a finite-dimensional system, the decomposition $\{p_i,\varphi_i\}$ minimizing the average variance of an operator $H^{(I)}$ in the state
\begin{equation}\label{eq:decomposition}
\rho_\omega=\sum_i p_i|\varphi_i\rangle\langle\varphi_i|
\end{equation}
is obtained from the solution of the Lyapunov equation
\begin{equation}\label{eq:Lyapunov}
\frac{1}{2}\{\rho_\omega,H^{(I)}_{(\omega)}\} = \rho_\omega^{1/2}H^{(I)}\rho_\omega^{1/2}.
\end{equation}
If $E_i$ and $|\Psi_i\rangle$ denote the eigenvalues and eigenvectors of $H^{(I)}_{(\omega)}$, then $p_i=\langle\Psi_i|\rho_\omega|\Psi_i\rangle$, and $p_i|\varphi_i\rangle\langle\varphi_i|=\rho_\omega^{1/2}|\Psi_i\rangle\langle\Psi_i|\rho_\omega^{1/2}$~\cite{Yu2013Quantum}. The corresponding minimum averaged variance is one quarter of the quantum Fisher information $F_Q(\rho_\omega;H^{(I)})$.

Let $\mathcal M_H$ denote the subset of full-rank density matrices for which $H^{(I)}_{(\omega)}$ has simple spectrum. On this regular set, the optimal decomposition is unique up to relabelling. Moreover, on a fixed candidate leaf, the optimality equations imply that the spectrum of $H^{(I)}_{(\omega)}$ is independent of the populations. Hence every state on the leaf determines the same optimal set $\{\varphi_i\}$.  This construction defines a foliation of $\mathcal M_H$\footnote{In a neighborhood of a state in $\mathcal M_H$, the populations $p_i$ provide $d-1$ coordinates along the leaf. The transverse coordinates can be chosen as the independent off-diagonal matrix elements, in the energy eigenbasis, of the symmetric logarithmic derivative associated with the Hamiltonian flow. These matrix elements encode the energy-coherence structure fixed by the leaf. In these adapted coordinates, a leaf is obtained by keeping the symmetric-logarithmic-derivative data fixed~\cite{Fagotti2026Quantum} and varying only the populations. It is therefore locally diffeomorphic to the open probability simplex, and its tangent space is spanned by the population directions. Smoothness of the tangent distribution follows from the smooth dependence of the adapted coordinates on the state, while involutivity is immediate because the population directions are coordinate vector fields.}---Figure~\ref{f:foliation} right (see also Ref.~\cite{Uhlmann2010Roofs}). 
From a qualitative physical perspective, a foliation based on dynamics and containing the stationary states as one of its leaves is the natural first step towards extending the equilibrium organizational scheme away from equilibrium. The minimum-variance foliation is distinguished by its intrinsic kinematic structure in state space: \emph{time translations generated by $H$ map leaves into leaves and preserve both the leaf coordinates and the Bures speed---the quantum speed---associated with $H$~\cite{BraunsteinCaves1994,Taddei2013QuantumSpeedLimit}}. 
This property led us to ask whether every leaf containing at least one regular state might exhibit the same kind of thermodynamic regularity as the stationary-state sector underlying equilibrium thermodynamics. Motivated by this possibility, in Ref.~\cite{Fagotti2026Quantum} we proposed that the eigenstate thermalization hypothesis might admit an analogue for regular states on each leaf, which we called ``leaf typicality''. In its strongest form, leaf typicality would imply that, for local observables, $\rho_{\omega}$ can be replaced by any pure state $|\varphi_i\rangle$ in the decomposition having the same energy. Preliminary numerical evidence was encouraging. The finite-volume construction, however, relies strongly on the finite dimensionality of the state space and does not by itself provide a basis for identifying an appropriate nonequilibrium generalization of temperature. Still, Ref.~\cite{Fagotti2026Quantum}  made an important first step in that direction by defining  leaf canonical ensembles in the same way as  canonical ensembles can be defined at equilibrium. Specifically, given the optimal decomposition \eqref{eq:decomposition}, the thermodynamic entropy was identified with the Shannon entropy of the population
\begin{equation}\label{eq:thentropy}
S_{\rm th}=-\sum_i p_i\log p_i\, ,
\end{equation} 
and the \emph{leaf canonical ensemble} was defined as the state maximizing the thermodynamic entropy under the constraints of normalization and energy~\cite{Jaynes1957Information,Jaynes1957Information2}. The result takes the form
\begin{equation}\label{eq:leafGibbs}
\frac{\rho_{\omega_\bullet}^{\frac{1}{2}}e^{-\beta H_{(\omega_\bullet)}^{(I)}}\rho_{\omega_\bullet}^{\frac{1}{2}}}{\mathrm{tr}[\rho_{\omega_\bullet} e^{-\beta H_{(\omega_\bullet)}^{(I)}}]} \, ,
\end{equation}
where $\rho_{\omega_\bullet}$ is the density matrix of the \emph{barycenter} of the leaf, which is the unique state on the leaf with  uniform population\footnote{The square-root dressing by the barycenter state is formally reminiscent of Scrooge ensembles, where a distinguished ensemble associated with a density matrix is obtained by applying the square root of that density matrix to an unbiased distribution of pure states \cite{JozsaRobbWootters1994}. In the present case, the dressing fixes the leaf, while the non-uniform weighting takes Gibbs form with respect to the $\omega$-dressed Hamiltonian. }.
This readily follows from the properties highlighted after \eqref{eq:Lyapunov}.  The strong analogy with the Gibbs ensemble naturally points to  identifying the inverse temperature as the derivative of the thermodynamic entropy with respect to the energy, i.e., $\beta$ in \eqref{eq:leafGibbs}.  There is, however, a subtlety that is both conceptual and operational: the definition \eqref{eq:leafGibbs} is anchored to the barycenter $\omega_\bullet$ of the leaf, which is a state with obscure physical properties. 

This exposes the central issue addressed in the present work: which part of the finite-volume construction survives as an intrinsic thermodynamic structure in the thermodynamic limit. 
In particular, neither the density matrix nor the optimal pure-state decomposition has a direct infinite-volume analogue. As discussed in the next sections, what survives in an infinite chain is the local, algebraic content of the construction. When the leaf-typicality hypothesis  is lifted to the thermodynamic level, the infinite-volume leaf becomes an equivalence class of regular thermodynamic states, while the population coordinates of the finite system are replaced by a single thermodynamic coordinate---see Figure~\ref{f:commutant} right. The important point is that this coordinate remains anchored to the post-quench Hamiltonian. It therefore retains an intrinsic physical meaning even out of equilibrium, unlike the coordinate associated with the commutant foliation, whose thermodynamic meaning was tied to the pre-quench equilibrium dynamics.

\subsection{First steps towards the thermodynamic limit}\label{s:leaves}

This section collects the finite-volume identities and their local thermodynamic-limit counterparts that will be used throughout the paper. Readers interested mainly in the thermodynamic construction may focus on the definitions of the $\omega$-dressed Hamiltonian, Eqs.~\eqref{eq:hw-def} and \eqref{eq:omegadressedH}, of the harmonic conjugate of $H$, Eqs.~\eqref{eq:hpm} and \eqref{eq:tildeHomega}, and on the local leaf condition~(B), Eq.~\eqref{eq:leaf-cond-2}, consulting the remaining details as needed.

We start with a clarification. Although the minimum-variance foliation is motivated by the problem of assigning temperature after a global quench, most of the paper\footnote{Time evolution re-enters in an essential way when we discuss subsystems.} does not refer explicitly to time evolution. The state under consideration should therefore be understood as an arbitrary regular state, for instance the state reached at any finite time after the quench.

We write $\omega'\sim_{\mathcal L_H}\omega $ when $\omega'$ and $\omega$ belong to the same leaf associated with the Hamiltonian $H$, and denote the corresponding leaf by
\begin{equation}
\mathcal L_H(\omega) =\{\omega' : \omega'\sim_{\mathcal L_H}\omega\}\, .
\end{equation}

In finite volume, a full-rank state $\rho_{\omega'}$ of an isolated system with Hamiltonian $H^{(I)}$ belongs to the leaf through $\rho_\omega$ if and only if
\begin{equation}\label{eq:leaf-commuting}
\bigl[\rho_\omega^{-1/2}\rho_{\omega'}\rho_\omega^{-1/2},\,H^{(I)}_{(\omega)}\bigr]=0\, .
\end{equation}
There is an equivalent dynamical characterization. Let the state be infinitesimally evolved by the Hamiltonian, $\dot\rho_\omega=i[\rho_\omega,H^{(I)}]$,  and define the symmetric logarithmic time derivative $L_{\rho_\omega}[\dot\rho_\omega]$, or SLtD, by
\begin{equation}\label{eq:SLD0}
\dot\rho_\omega=\frac{1}{2}
\left\{
	\rho_\omega,L_{\rho_\omega}[\dot\rho_\omega]
\right\}.
\end{equation}
Then two full-rank states belong to the same leaf if and only if they have the same SLtD. Thus the leaf can be characterized either algebraically, by \eqref{eq:leaf-commuting}, or dynamically, by the infinitesimal motion generated by the Hamiltonian. In the following we use these finite-volume characterizations introduced in Ref.~\cite{Fagotti2026Quantum}  to formulate the relation ``being on the same leaf'' in a way that remains meaningful in the thermodynamic limit.

The first key observation is that, while \eqref{eq:Lyapunov} is not directly available in the thermodynamic limit, its solution can be cast in a form that can also be used in infinite chains. To that aim, we define the \emph{$\omega$-dressed density} associated with $h_\ell$ by
\begin{equation}\label{eq:hw-def}
h_{(\omega), \ell}^{(I)}(\lambda)=\int_{-\infty}^{\infty}\frac{ds}{\cosh(\pi s)}\, \sigma_0^{(s-i\lambda)\beta_0}(h_\ell)\, ,
\end{equation}
where the dependence on $I$ enters only through the finite-volume dynamics $\sigma_0^t$ and is exponentially small in the distance of $\ell$ from the boundary of $I$. In the following the superscript $(I)$ will often be omitted, $h_{(\omega), \ell}(\lambda)$ denoting either  $h_{(\omega), \ell}^{(I)}(\lambda)$ or its limit $\lim_{I\nearrow\mathbb Z}h_{(\omega), \ell}^{(I)}(\lambda)$, depending on the context.
For one-dimensional quantum spin chains, complex-time analyticity of local observables is a standard consequence of locality estimates: in the finite-range case this goes back to Araki~\cite{Araki1969Gibbs}, while exponentially decaying interactions are addressed by the more recent extension of Ref.~\cite{PerezGarciaPerezHernandez2023}, with the caveat that the width of the analyticity strip depends on the interaction’s decay. In particular, for local $h_{0,\ell}$ the analytic continuation $\sigma_0^{\beta_0(s-i\lambda)}(h_\ell)$ exists for $|\lambda|\le \frac{1}{2}$. We may therefore define the boundary values
\begin{equation}\label{eq:hpm}
h_{(\omega),\ell}^{\pm}=\lim_{\lambda\to \pm\frac{1}{2}} h_{(\omega),\ell}(\lambda), \qquad h_{(\omega),\ell}^{-}=(h_{(\omega),\ell}^{+})^\dagger.
\end{equation}
We also use the shorthand
\begin{equation}
h_{(\omega),\ell}=h_{(\omega),\ell}(0).
\end{equation}

One can readily check in a finite chain that $h_{(\omega),\ell}$ is the density of $H^{(I)}_{(\omega)}$, i.e.,  
\begin{equation}\label{eq:omegadressedH}
H_{(\omega)}^{(I)}=\sum_{\ell\in I} h_{(\omega),\ell}\, .
\end{equation} 
We also point out
\begin{equation}
\begin{aligned}
h_\ell=&\frac{1}{2}(h^{-}_{(\omega),\ell}+h^{+}_{(\omega),\ell})\\
H^{(I)}=&\frac{1}{2}(H^{(I)-}_{[\omega]}+H^{(I)+}_{[\omega]})\, ,
\end{aligned}
\end{equation}
where $H^{(I)\pm}_{[\omega]}$ denote the finite-volume operators with density $h^{\pm}_{(\omega),\ell}$. A second useful quantity to define in finite chains is the  ``harmonic conjugate'' of $H$
\begin{equation}\label{eq:tildeHomega}
\tilde H^{(I)}_{[\omega]}=\frac{i}{2}(H^{(I)-}_{[\omega]}-H^{(I)+}_{[\omega]}) 
\end{equation}
and the associated density 
\begin{equation}\label{eq:harmonicdensity}
\tilde h_{(\omega),\ell}=\frac{i}{2}(h^{-}_{(\omega),\ell}-h^{+}_{(\omega),\ell})\, .
\end{equation}
This is directly related to the symmetric logarithmic time derivative $L_{\rho_\omega}[\dot \rho_\omega]$
\begin{equation}
\tilde H^{(I)}_{[\omega]}=\frac{1}{2} L_{\rho_\omega}[\dot \rho_\omega]\, ;
\end{equation}
indeed it satisfies the equation
\begin{equation}\label{eq:SLD}
\{\rho_{\omega},\tilde H^{(I)}_{[\omega]}\}=i[\rho_{\omega},H^{(I)}]
\end{equation}
and, since the system is isolated, the right hand side is the time derivative of the density matrix. 
We would like to clarify that \emph{we have used the notation $\tilde H_{[\omega]}$ instead of $\tilde H_{(\omega)}$ to emphasize that every state $\omega'\in\mathcal L_H(\omega)$ on the leaf across $\omega$ is associated} with the same SLtD, i.e., \emph{with the same $\tilde H_{[\omega]}$}; yet, we have distinguished between $\tilde H^{(I)}_{[\omega]}$ and $L_{\rho_\omega}[\dot \rho_\omega]$, since the two quantities are directly related only when the dynamics are unitary, which is lost when considering subsystems. A similar comment applies to $H^\pm_{[\omega]}$.  

This leads to a particularly transparent formulation of the leaf condition in the thermodynamic limit

\begin{description}

\item[Leaf condition (A)] 
\begin{equation}\label{eq:leaf-cond-SLD}
\omega'\sim_{\mathcal L_H} \omega \quad\Longleftrightarrow\quad \tilde h_{(\omega)}\sim_{\Sigma} \tilde h_{(\omega')}\, ,
\end{equation}
where we used the shorthand $q\equiv \{q_\ell\}_{\ell\in \mathbb Z}$ and $\sim_{\Sigma}$ means that finite sums of the densities in the corresponding sets agree up to boundary terms.  Note that, while $\tilde H_{[\omega]}$ is a property of the leaf, its density $\tilde h_{(\omega),\ell}$ depends on the particular state $\omega$ on the leaf.

We stress that the harmonic conjugate $ \tilde H_{[\omega]}$ characterizes the leaf only if the Hamiltonian $H$ is understood. Alternatively, we can collect the two operators together into the pseudo-Hermitian operator $H_{[\omega]}^-$ with quasilocal density $h^{-}_{(\omega)}=h-i\tilde h_{(\omega)}$\footnote{ Although we will not explore this connection further here, the very existence of this structure points to a close relation with pseudo-Hermitian statistical mechanics~\cite{Jakubsky2007THERMODYNAMICS,Mostafazadeh2010PSEUDO,Cao2023Statistical}, where the state would naturally play the role of the metric.}.

\end{description}

While \eqref{eq:leaf-cond-SLD} is conceptually strong, there is an alternative formulation of the leaf condition that has a more direct operational value.  

\begin{description}

\item[Leaf condition (B)] for every local Hermitian observable $O$,
\begin{equation}\label{eq:leaf-cond-2}
\begin{gathered}
\omega'\in \mathcal L_H(\omega)\Leftrightarrow  \omega'(H_{[\omega]}^{(I)+} O-O\,H_{[\omega]}^{(I)-})\xrightarrow{I\nearrow\mathbb Z}0\\
H_{[\omega]}^{(I)\pm}=\sum_{\ell\in I} h^{\pm}_{(\omega),\ell}=\sum_{\ell\in I} h_\ell\pm i \tilde h_{(\omega),\ell}\, ,
\end{gathered}
\end{equation}
that is to say, $\omega'\in \mathcal L_H(\omega)$ iff $\omega'( \tilde H^{(I)}_{[\omega]})\xrightarrow{I\nearrow\mathbb Z}0$ and
\begin{equation}\label{eq:leaf-cond-3}
\langle\langle \tilde H^{(I)}_{[\omega]},O\rangle\rangle_{\omega'}^c-\tfrac{i}{2}\omega'([H^{(I)},O])\xrightarrow{I\nearrow\mathbb Z}0\, .
\end{equation}
For the moment, and until the end of this section, we leave aside the question of the existence of the limit $I\nearrow\mathbb Z$. We stress, however, that only a subset of states should be expected to admit a proper thermodynamic-limit description.

We can readily show the equivalence with \eqref{eq:leaf-commuting} in a finite system. First, we multiply \eqref{eq:leaf-commuting} by $\rho_\omega^{1/2}$ both on the right and on the left hand side, obtaining the  intertwining relation
\begin{equation}\label{eq:leaf-intertwine}
H^{(I)-}_{[\omega]} \rho_{\omega'}=\rho_{\omega'}H^{(I)+}_{[\omega]}\, ,
\end{equation}
where we used $H^{(I)\pm}_{[\omega]}=\rho_\omega^{\mp 1/2}H^{(I)}_{(\omega)}\rho_\omega^{\pm 1/2}$.  Then, we multiply \eqref{eq:leaf-intertwine} by a generic operator $O$ and take the expectation value. Imposing the resulting conditions for a complete set of operators gives back \eqref{eq:leaf-intertwine}. In the thermodynamic limit the conditions can be imposed only for local operators, that is to say, \eqref{eq:leaf-cond-2}. This condition should be compared with the stationarity one appearing in the definition of equilibrium states, which can be expressed as $\lim_{I\nearrow \mathbb Z}\omega([H^{(I)},O])=0$, for all local $O$. 

\end{description}

The minimum-variance leaves are obtained by holding the full energy-coherence structure fixed, but the amount of coherent energy fluctuations depends on the state and is quantified by its quantum Fisher information with respect to the Hamiltonian. Appendix~\ref{ss:QFI} clarifies what remains, in the thermodynamic limit, of the connection between the leaves and the QFI. Here we only highlight that the finite-volume QFI can be decomposed into a sum of local densities, each of which admits a thermodynamic limit. Specifically, a local density of the QFI with respect to the Hamiltonian is
\begin{equation}\label{eq:QFI}
\mathcal F_Q^{(\omega')}(\omega;h_\ell)=2\,\omega\!\left(i[H,\tilde h_{(\omega'),\ell}]\right)\, ,
\end{equation}
where $\omega'$ can be any state in $\mathcal L_H(\omega)$. Different choices of $\omega'$ correspond to different density representatives of the same extensive QFI.

\section{Nonequilibrium temperature in the thermodynamic limit}\label{s:isolated}

The goal of this section is to show that, at fixed energy coherence, the thermodynamic limit generically selects a one-dimensional variety of regular states. This will then allow us to  attach an inverse temperature $\beta_\omega^H$ to the state $\omega$ of an isolated system that was prepared in a Gibbs state (i.e., in a KMS state) and was then put out of equilibrium by quenching the interaction into $h$. 

As discussed in Section~\ref{ss:min-var}, the intuitive definition of inverse temperature that follows from the leaf-canonical ensemble~\eqref{eq:leafGibbs} is undermined by the lack of a clear physical interpretation of the barycenter $\omega_\bullet$ of the leaf. On the other hand, the leaf is specified by a state, $\omega$, with a temperature that is a priori unknown, but with physical properties, such as clustering, that are transparent. It becomes more convenient to generalize a derivation of Gibbs ensembles based on the principle of minimum discrimination information~\cite{Shore1980Axiomatic,Kullback1951OnIA}, rather than the derivation of Section~\ref{ss:min-var}, which is based on the principle of maximum entropy. Specifically, imagine starting from a canonical ensemble $\rho^{Gibbs}_{\beta_\ast}$ with unknown temperature $\beta_\ast$. It is simple to show that any other canonical ensemble can be obtained by \emph{minimizing the relative entropy of the population} $\{p\}$ with respect to that, let us call it $\{q\}$, of $\rho^{Gibbs}_{\beta_\ast}$
\begin{equation}
D_{KL}(\{p\}|\{q\})=\sum_i p_i\log \frac{p_i}{q_i}\, ,
\end{equation} 
under normalization and fixed mean energy. The result of the variation  is indeed 
\begin{equation}\label{eq:Gibbs1}
\rho^{Gibbs}_{\Delta\beta|\beta^\ast}=\frac{\rho^{Gibbs}_{\beta_\ast} e^{-\Delta\beta H^{(I)}}}{\mathrm{tr}(\rho^{Gibbs}_{\beta^\ast}e^{-\Delta\beta H^{(I)}})}\, .
\end{equation}
The drawback is that this approach gives a relative inverse temperature rather than an absolute one, requiring, in turn, a reference point. In that respect,  the state with zero inverse temperature is the tracial state $\tau$, which can be characterized, within the family of Gibbs states, by its invariance under a generic rescaling of time (i.e., of $H^{(I)}$). Specifically, it is sufficient to impose
\begin{equation}\label{eq:equationGibbs}
\frac{\mathrm{tr}(\rho^{Gibbs}_{\Delta\beta|\beta^\ast}H^{(I)})-\tau(H^{(I)})}{|I|}=0
\end{equation}
(the subtraction makes the condition independent of the zero of energy), which, the energy being strictly monotonic in $\Delta\beta$, is solved within the Gibbs family by the tracial state alone.
The temperature $\beta_\ast$ can then be extracted by imposing  $\rho^{Gibbs}_{\Delta\beta|\beta^\ast}$ to be the infinite-temperature state, which results in $\beta_\ast=-\Delta\beta$.  

We now apply the same logic to the leaf $\mathcal L_H(\omega)$ across $\omega$. The minimization of the relative entropy of the population gives
\begin{equation}\label{eq:rhoomegadef}
\frac{\rho_\omega^{\frac{1}{2}}e^{-\Delta \beta H^{(I)}_{(\omega)}}\rho_\omega^{\frac{1}{2}}}{\mathrm{tr}(\rho_\omega^{\frac{1}{2}}e^{-\Delta \beta H^{(I)}_{(\omega)}}\rho_\omega^{\frac{1}{2}})}\, ,
\end{equation}
which is to  \eqref{eq:Gibbs1} what \eqref{eq:leafGibbs} is to the Gibbs ensemble. However, the physical properties of \eqref{eq:rhoomegadef} for $\Delta\beta\neq 0$ do not follow directly from the finite-volume construction, hence it is not evident that the analogue of \eqref{eq:equationGibbs} would make physical sense in the thermodynamic limit. On the other hand, since $\omega$ is a KMS state of a spin chain, it has exponentially decaying connected correlations; see also Ref.~\cite{FrohlichUeltschi2015Correlations} for high-temperature results on uniqueness of the thermodynamic-limit KMS state and exponential decay of equilibrium correlations in broad classes of quantum lattice systems. 

Appendix~\ref{a:leafthrough} provides a detailed argument that, within the regularity regime considered there, the finite-volume functionals corresponding to \eqref{eq:rhoomegadef},
\begin{equation}\label{eq:omegadef0}
O\mapsto F_{O}^{(I)}(\Delta \beta)=
\frac{
	\omega\!\left(e^{-\frac{\Delta \beta}{2} H_{[\omega]}^{(I)+}}Oe^{-\frac{\Delta \beta}{2} H_{[\omega]}^{(I)-}}\right)
}{
	\omega\!\left(e^{-\frac{\Delta \beta}{2} H_{[\omega]}^{(I)+}}e^{-\frac{\Delta \beta}{2} H_{[\omega]}^{(I)-}}\right)
}\, ,
\end{equation}
converge locally uniformly as $I\nearrow \mathbb Z$ throughout an interval $|\Delta\beta|<\Delta\beta_*$, where $\Delta\beta_*>0$ can be chosen independently of $I$. Their thermodynamic limit is analytic in $\Delta\beta$ and, for real $\Delta\beta$, defines an exponentially clustering state. 
Specifically, we adapt the high-temperature cluster-expansion results of Ref.~\cite{NguyenFernandez2024HighTemperature} to the ordered product appearing in \eqref{eq:omegadef0}\footnote{We do not  cast this argument as a complete rigorous proof of the thermodynamic-limit construction: doing so would require a fully $C^\star$-algebraic formulation of the theory in terms of quasilocal observables, which lies beyond the scope of this work.}.
Within this regularity regime, the relative nonequilibrium inverse temperature has an intrinsic meaning: it measures the change in the canonical-flow coordinate from $\omega$ to another state in a one-parameter family of exponentially clustering thermodynamic states.

The remaining step is to identify an anchor that allows for a definition of  absolute inverse temperature.   Having in mind the invariance under a generic rescaling of time, which identifies the infinite-temperature state at equilibrium, we borrow the same condition, \eqref{eq:equationGibbs}, and enforce that the infinite-temperature state has the same energy density as the tracial state.
Thus, if the negative of the (still unknown) inverse temperature $\beta_\omega^H$ of $\omega$ lies in the domain of analyticity of $F_{O}^{(I)}(\Delta \beta)$ (which is the case at sufficiently high pre-quench temperature; see the end of Appendix~\ref{a:leafthrough}), then it can be obtained as the solution to the equation
\begin{equation}\label{eq:condT0}
\lim_{I\nearrow \mathbb Z}\frac{F^{(I)}_{H^{(I)}}(-\beta_\omega^H)-\tau(H^{(I)})}{|I|}=0\, .
\end{equation}
Note that, in finite volume, Eq.~\eqref{eq:condT0} admits a unique solution. Indeed, the energy $F_{H^{(I)}}^{(I)}(\Delta \beta)$ is strictly decreasing in $\Delta\beta$---see Eq.~\eqref{eq:partomega} in Section~\ref{s:generic}---and spans the open interval $(\min_i E_i,\max_i E_i)\ni \tau(H^{(I)})$.  Appendix~\ref{a:leafthrough} shows that the existence of the solution of \eqref{eq:condT0} carries over in the thermodynamic limit, at least for sufficiently high pre-quench temperatures.  We remark that $\tau(H^{(I)})$ is also the finite-volume energy of the barycenter of the leaf; Eq.~\eqref{eq:condT0} can therefore be read as enforcing a matching that, under leaf typicality, amounts to \emph{local equivalence between the reference state and the infinite-temperature state of the leaf canonical ensemble} as $|I|\rightarrow \infty$. 

We are then in a position to define the leaf canonical ensemble $\omega_\beta$ through $\omega$ at inverse temperature $\beta$ as follows
\begin{equation}\label{eq:omegabeta0}
\omega_{\beta}: O\mapsto \lim_{I\nearrow \mathbb Z} F_{O}^{(I)}(\beta-\beta_\omega^H)\, .
\end{equation}
In the following we will call $(\omega_{\beta_0},\Delta \beta)\mapsto \omega_{\beta_0+\Delta \beta}$ the canonical flow. 
We note a slight abuse of terminology, adopted for simplicity: we continue to use the term ``leaf canonical ensemble'' also for $\omega_{\beta}$, adding the qualification ``through $\omega$''. Accordingly, the original leaf canonical ensemble~\eqref{eq:leafGibbs} may be referred to as the leaf canonical ensemble through the barycenter\footnote{This terminology is motivated by a Bayesian statistical interpretation of the two constructions, in which the barycenter corresponds to a uniform prior over the leaf populations, whereas the construction through $\omega$ uses the populations of the regular physical state $\omega$ as the reference prior. }.

Finally, the argument given above is restricted to a finite interval of inverse temperatures. By analogy with equilibrium quantum spin chains, however, it is natural to expect the leaf-canonical flow to remain regular for all finite real inverse temperatures. 
A breakdown of such regularity could signal a phase transition on a noncommuting leaf, a remarkable possibility that deserves separate investigation.

\subsection{Generic leaves}\label{s:generic}

Having established that the canonical flow defines a thermodynamic-limit state $\omega_\beta$ within the regularity regime discussed above, we can identify the infinitesimal generator of the flow with the action of a \emph{pseudolocal charge} on the leaf, in a sense analogous to that of Ref.~\cite{Doyon2017Thermalization}. Indeed, differentiating \eqref{eq:omegadef0} at $\Delta\beta=0$ gives $\left.\partial_{\Delta\beta}F_O(\Delta\beta)\right|_{\Delta\beta=0}=-\widehat H_{\omega}(O)$, with
\begin{equation}\label{eq:canonical}
\begin{aligned}
\widehat H_{\omega}(O)=&
\lim_{I\nearrow\mathbb Z}
\left[
	\langle\!\langle H^{(I)},O\rangle\!\rangle_{\omega}^{c}+\frac{i}{2}\omega\!\left([\tilde H^{(I)}_{[\omega]},O]\right)
\right]\\
\equiv & \frac{1}{2}\lim_{I\nearrow\mathbb Z}
\left[
	\langle H^{(I)+}_{[\omega]},O\rangle^c_{\omega}+\langle O,H^{(I)-}_{[\omega]}\rangle^c_{\omega}
\right]\, .
\end{aligned}
\end{equation}
In this formulation $H^{(I)+}_{[\omega]}$ and $H^{(I)-}_{[\omega]}$ can be regarded as the left and right pseudolocal sequences associated with
the charge. Incidentally, if we specialize  \eqref{eq:canonical} to the Hamiltonian, we find\footnote{Using the optimal decomposition \eqref{eq:decomposition} for $\omega_\beta$, with population $\{p_{\beta,i}\}$, we also have $\partial_{\beta}\omega_\beta(H^{(I)})=-\sum_i p_{\beta,i}(\langle\varphi_i|H^{(I)}|\varphi_i\rangle-\omega_\beta(H^{(I)}))^2$.}
\begin{equation}\label{eq:partomega}
\partial_{\beta}\omega_\beta(H^{(I)})=-\langle H^{(I)},H^{(I)}\rangle^c_{\omega_\beta}+\frac{1}{4}F_Q(\omega_\beta,H^{(I)})\, ,
\end{equation}
which is strictly negative, implying that \eqref{eq:condT0} has a unique solution. We exploit this identity to clarify in which sense the states $\omega_\beta$ are distinguished from the other states of the leaf (in finite volume), even though leaf typicality is a statement of local equivalence.  While leaf typicality implies that the last term of \eqref{eq:partomega}, i.e., the QFI with respect to the Hamiltonian, is invariant under an isoenergetic change of the state on the leaf, it does not imply anything concerning the energy variance (e.g. in the commuting leaf, the energy variance in a Gibbs state is extensive, whereas it vanishes in every excited state of the Hamiltonian). Thus, the incoherent fluctuations appearing on the right hand side of \eqref{eq:partomega} and, in turn, the inverse temperature $\beta$ have a thermodynamic meaning only along the orbit of the canonical flow. 

A question that has remained open is what dimensionality should characterize the variety of thermodynamically regular states on a leaf. A naive parameter count offers some guidance. In finite volume, both the leaves of the commutant foliation---Section~\ref{s:mVfoliation}---and those of the minimum-variance foliation are $(d-1)$-dimensional, while the state space has dimension $d^2-1$. On a commutant leaf containing a thermal state of a nonintegrable quasilocal Hamiltonian, the eigenstate thermalization hypothesis reduces, as far as local observables are concerned, the $(d-1)$-dimensional family to a one-dimensional thermodynamic variety. If this reduction persists generically across the thermodynamically relevant subset of commutant leaves, dimensional counting suggests that a transverse $(d-1)$-dimensional leaf of another foliation should likewise retain a single thermodynamic direction through the thermal state, i.e., intersect the thermodynamically regular states in a one-dimensional variety. The canonical flow constructed above provides precisely such a one-dimensional regular family on the minimum-variance leaf. This motivates the following definition.
\begin{description}

\item[Genericity] We shall call the leaf \emph{generic around $\omega$} if the canonical charge $\widehat H_\omega$ is the unique  nontrivial pseudolocal charge on the leaf in a neighborhood of~$\omega$, up to rescaling.  We call the leaf \emph{generic} if it is generic around every regular state in $\mathcal L_H(\omega)$.

\end{description}

Generic leaves are the nonequilibrium counterparts of the commuting sector of a generic Hamiltonian, namely of a dynamics with no thermodynamically relevant conservation laws beyond the energy and no additional exceptional structure. 
On generic leaves, the leaf typicality hypothesis proposed in Ref.~\cite{Fagotti2026Quantum} is expected to have the same status as the eigenstate thermalization hypothesis for generic equilibrium dynamics. This analogy leads us to reconsider the relation between the leaf-canonical ensemble~\eqref{eq:leafGibbs} and the regular states generated by the canonical pseudolocal flow~\eqref{eq:omegabeta0},  that is to say, the relation between the principle of maximum entropy (in which the leaf is treated as a bare probability simplex with no distinguished state) and the principle of minimum discrimination information. In equilibrium the two principles are equivalent, and it is not evident whether replacing maximum entropy by minimum discrimination information out of equilibrium is merely a technical expedient or reflects a genuine physical consequence of combining the minimum-variance foliation with locality. The deformation of the orbit with respect to the path associated with the leaf canonical ensemble can be traced back to  the bias of the  inverse-temperature estimator $ \beta^{(I)}_{\mathrm{bare}}$ given by
\begin{equation}\label{eq:dressing}
\beta^{(I)}_{\mathrm{bare}}(\beta)=\frac{\partial_\beta S^{(I)}_{\mathrm{th}}(\omega_\beta|\mathcal L_{H^{(I)}}(\omega))}{\partial_\beta \omega_\beta(H^{(I)})}\, ,
\end{equation} 
which would approach $\beta$ in the limit $I \nearrow \mathbb Z$ under the principle of maximum entropy.  Here and henceforth, $S_{\mathrm{th}}(\omega'|\mathcal L_H(\omega))$, for $\omega'\in\mathcal L_H(\omega)$, denotes the thermodynamic entropy of $\omega'$ computed from the population coordinates of the specified leaf (making the leaf dependence explicit will become essential in Section~\ref{s:subsystems}). In  Section~\ref{s:examples}  we will provide numerical evidence that $\lim_{I \nearrow \mathbb Z}\beta^{(I)}_{\mathrm{bare}}$ is generically biased. 

Finally, we cannot exclude that, even when the leaf admits only one pseudolocal thermodynamic direction locally, the regular thermodynamic states on the leaf are not all connected by admissible pseudolocal flows. 

\begin{description}

\item[Fragmentation]
We shall call a leaf \emph{thermodynamically fragmented} if the regular states on the leaf decompose into more than one thermodynamic sector, namely into distinct classes of states that cannot be connected to one another by admissible pseudolocal flows.

\end{description}

We leave the analysis of thermodynamically fragmented leaves, and of their possible relation to Hilbert-space fragmentation or phase transitions, for future work. 

\subsection{Non-generic leaves}\label{ss:nongeneric}

Returning to finite volume, the canonical flow describes only a one-dimensional variety of states on the leaf, and it is natural to wonder whether other flows could exist connecting $\omega$ with other regular states.  From \eqref{eq:leaf-commuting} it readily follows that such potential flows should have the form 
\begin{equation}\label{eq:omegadef1}
F_{O,Q}^{(I)}(\Delta \mu)=
\frac{
	\omega(e^{-\frac{\Delta \mu}{2} Q_{[\omega]}^{(I)+}}Oe^{-\frac{\Delta \mu}{2} Q_{[\omega]}^{(I)-}})
}
{
	\omega(e^{-\frac{\Delta \mu}{2} Q_{[\omega]}^{(I)+}}e^{-\frac{\Delta \mu}{2} Q_{[\omega]}^{(I)-}})
}
\end{equation}
for some $Q_{[\omega]}^{(I)-}$ associated with a $\omega$-dressed operator commuting with $H_{(\omega)}^{(I)}$, $[Q_{(\omega)}^{(I)},H_{(\omega)}^{(I)}]=0$. The notation $Q_{[\omega]}^{(I)\pm}$ emphasizes that these left and right representatives are properties of the leaf, rather than of a particular state on it\footnote{Indeed, since $H_{(\omega)}^{(I)}$ is assumed to have simple spectrum, every operator commuting with it is diagonal in the same basis. If, in addition, $Q_{(\omega)}^{(I)}$ has full rank and simple spectrum, it fixes this common diagonalizing basis, up to phases and permutations. Consequently, $\frac{1}{2}(Q_{[\omega]}^{(I)+}+Q_{[\omega]}^{(I)-})$ has the same optimal decomposition as $H^{(I)}$. This is therefore a property of the whole leaf, not of the particular representative state $\omega$.}.

If $Q_{[\omega]}^{(I)\pm}$ have the same locality properties as $H_{[\omega]}^{(I)\pm}$, we can use the same argument as for the canonical flow and conclude that the sequence $Q_{(\omega)}^{(I)}$ defines an independent pseudolocal charge on the leaf
\begin{equation}
\begin{gathered}\label{eq:charge1}
\widehat Q_{\omega}(O)=
\lim_{I\nearrow\mathbb Z}
\left[
	\langle\!\langle \bar Q_{[\omega]}^{(I)},O\rangle\!\rangle_{\omega}^{c}+\frac{i}{2}\omega\!\left([\tilde Q^{(I)}_{[\omega]},O]\right)
\right]\\ ([Q_{(\omega)}^{(I)},H^{(I)}_{(\omega)}]=0)\, ,
\end{gathered}
\end{equation}
where $\bar Q_{[\omega]}^{(I)}=\frac{1}{2}(Q_{[\omega]}^{(I)+}+Q_{[\omega]}^{(I)-})$ and $\tilde Q_{[\omega]}^{(I)}=\frac{i}{2}(Q_{[\omega]}^{(I)-}-Q_{[\omega]}^{(I)+})$---cf.~\eqref{eq:tildeHomega}. The deformation \eqref{eq:omegadef1} then gives rise to a leaf grand-canonical ensemble through $\omega$. Analogously, if $H_{(\omega)}^{(I)}$ is integrable and admits infinitely many independent pseudolocal conservation laws, the same construction leads to a leaf generalized Gibbs ensemble. More generally, we will call $\widehat Q_{\omega}$ a pseudolocal charge if $\widehat Q_{\omega}(O)$ connects regular states in a neighborhood of $\omega$, and call  $\lim_{I\nearrow\mathbb Z}(\omega(\bar Q_{[\omega]}^{(I)})/|I|)$ its integral of motion.

The presence of additional pseudolocal charges requires a better characterization of the reference state, i.e., the infinite-temperature state, with respect to which the inverse temperature and the additional Lagrange multipliers are computed. We propose the following
\begin{description}

\item[Infinite-temperature state ] The unique%
\footnote{For several independent leaf charges $\widehat Q_{j,\omega}$, the Jacobian relating the parameters of the corresponding pseudolocal flows to their integrals of motion is minus their Gram matrix with respect to the incoherent inner product~\eqref{eq:incoherentproduct},
$$
G_{j n}(\omega)=\lim_{I\nearrow\mathbb Z} \tfrac{\langle\langle Q_{j,(\omega)}^{(I)}, Q_{n,(\omega)}^{(I)}\rangle\rangle^c_\omega}{|I|}\, .
$$
Since the charges are linearly independent modulo the null space of the incoherent inner product, $G(\omega)$ is positive definite throughout the regular sector. The map from the pseudolocal-flow parameters to the integrals of motion is therefore locally invertible. On a convex domain of regular flow parameters, the same positivity also implies global injectivity, so that prescribed values of the integrals of motion identify at most one state in that domain.} 
thermodynamic state that is connected to $\omega$ by pseudolocal flows and has the same integrals of motion associated with a complete set of independent pseudolocal charges as the barycenter of $\mathcal L_H(\omega)$.

\end{description}
This definition allows one to identify thermal states also on non-generic leaves. As at equilibrium, their definition does not depend on whether additional conserved quantities are present: although a generic thermodynamic state may require additional coordinates, the thermal states form the canonical orbit through the infinite-temperature state. Under the corresponding generalized form of leaf typicality, the infinite-temperature state can equivalently be characterized as the unique thermodynamic state connected to $\omega$ by pseudolocal flows that is locally equivalent, as $I\nearrow\mathbb Z$, to the barycenter of $\mathcal L_H(\omega)$.

We point out that the additional pseudolocal charges are intrinsic to the leaf $\mathcal L_H(\omega)$. A generic displacement of $\omega$ in a direction transverse to the leaf destroys the additional commuting structure. A notable exception is provided by the time-translation automorphism: if $\omega$ is replaced by $\omega\circ\sigma^t$, the charge density is transported covariantly,
\begin{equation}
q_{(\omega\circ\sigma^t),\ell}^{(I)}=\sigma^{-t}\!\bigl(q_{(\omega),\ell}^{(I)}\bigr).
\end{equation}
Thus time evolution preserves the leaf charge structure. Nevertheless, by Lieb-Robinson bounds~\cite{LiebRobinson1972,BravyiHastingsVerstraete2006,NachtergaeleSims2010}, a density initially localized near $\ell$ can be approximated at time $t$ by an observable supported within a distance proportional to $|t|$, with exponentially small tails outside the corresponding light cone. Hence the effective range of $\bar Q_{[\omega\circ\sigma^t]}^{(I)}$ is expected to grow linearly with $|t|$: the charge becomes progressively less local as time increases.

The physical consequences of the additional pseudolocal flows are not immediately visible at the level of static correlations. Indeed, the reference state $\omega$ is assumed to be a standard KMS state and may therefore exhibit  ordinary local properties, independently of whether its leaf is generic. Rather, the structure of a leaf is expected to affect quantities that depend explicitly on the interaction and probe coherence properties of the state.

We leave this direction for future work and focus in the following on generic leaves.  This subset of $\mathcal M_H$ is expected to be dense in the space of regular thermodynamic states, since the existence of additional pseudolocal charges is an exceptional condition and is generically destroyed by arbitrarily small perturbations of the state, hence of the dressed Hamiltonian $H_{(\omega)}$.

\subsubsection{The incoherent inner product}\label{ss:inner}

Once additional pseudolocal flows are allowed, it is useful to equip the corresponding space of generators with an inner product. This provides the natural Hilbert-space structure needed, for instance, to define projections onto pseudolocal charges. On a leaf of the minimum-variance foliation, we find that the relevant pairing is a member---the harmonic mean---of the family of quantum covariances discussed in Ref.~\cite{Petzbook}. Specifically, we define
\begin{multline}\label{eq:incoherentproduct}
(A,B)^{\rm inc}_\omega =\\
\langle (A^\dagger)^+_{(\omega)},B\rangle^c_\omega
\equiv\langle A^\dagger,B^-_{(\omega)}\rangle^c_\omega
\equiv\langle\langle A^\dagger_{(\omega)},B_{(\omega)}\rangle\rangle^c_\omega\, .
\end{multline}
Here we use the notation $(A^\dagger)^+_{(\omega)}$ and $B^-_{(\omega)}$, rather than $(A^\dagger)^+_{[\omega]}$ and $B^-_{[\omega]}$, because the $\pm$ representatives of generic operators depend on the particular state $\omega$ (they depend on $\omega$ only through its leaf when the corresponding $\omega$-dressed operators commute with $H_{(\omega)}$). The second equality follows from the leaf condition~\eqref{eq:leaf-cond-2}, whereas the last one is a direct consequence of the Lyapunov equations \eqref{eq:Lyapunov} satisfied by the operators. Equivalently, this pairing identifies the generator of a pseudolocal flow with the corresponding linear functional on local observables: the flow generated by $A$ acts on $B$ as  (cf.~\eqref{eq:canonical}) $\widehat A_\omega(B)=(A,B)^{\rm inc}_\omega$. Note that the form $(\cdot,\cdot)^{\rm inc}_\omega$ is positive semidefinite and therefore defines an inner product after quotienting by its null space (which, in $\mathcal M_H$, consists of multiples of the identity). We call it the \emph{incoherent inner product}: for a self-adjoint observable $A$, the associated norm satisfies
\begin{equation}\label{eq:incoherentnorm}
\|A\|_{\omega,{\rm inc}}^2=
(A,A)^{\rm inc}_\omega=
\mathrm{Var}_\omega(A)-\frac{1}{4}F_Q(\omega,A)\, .
\end{equation}
Thus, the squared norm measures the part of the variance that is not accounted for by the coherent (QFI detectable) fluctuations. This suggests a possible route toward a fully $C^\star$-algebraic formulation of the theory. One could adapt the Hilbert-space construction of pseudolocal charges of Ref.~\cite{Doyon2017Thermalization} by replacing the susceptibility pairing with the incoherent inner product, while using the clustering results of Appendix~\ref{a:leafthrough} to establish the regularity of the corresponding flows. 

From \eqref{eq:omegabeta0}, \eqref{eq:partomega}, and \eqref{eq:condT0}, it then follows that the inverse temperature $\beta^H_\omega$ is the solution to the equation 
\begin{equation}\label{eq:betainner}
\lim_{I\nearrow\mathbb Z}\frac{1}{|I|}\Bigl[\omega(H^{(I)})-\tau(H^{(I)})+\int_0^{\beta^H_\omega} \|H^{(I)}\|_{\omega_\beta,{\rm inc}}^2 d\beta\Bigr]=0\, .
\end{equation}
This makes mathematically explicit the status of $\beta^H_\omega$ as the nonequilibrium generalization of inverse temperature. At equilibrium, the same relation determines $\beta$, the incoherent norm reducing to the Kubo–Mori–Bogoliubov norm of the Hamiltonian on the commuting leaf; away from equilibrium, the incoherent norm subtracts the coherent contribution to the energy fluctuations.

\subsection{On the meaning of nonequilibrium temperature}

The full system is isolated, and time evolution maps leaves into leaves while transporting the global energy--coherence structure covariantly. Consequently, the inverse temperature assigned to the full state is independent of time. At first sight, this may appear to conflict with the usual expectation in nonequilibrium many-body physics that temperature acquires dynamical significance through local equilibration and energy transport. The tension is only apparent: the global thermodynamic coordinate is conserved, whereas its dynamical role re-emerges upon restriction to subsystems---discussed in the next section---whose states and energy content evolve under transport. The inverse temperature of the full system therefore has no immediate counterpart in standard nonequilibrium descriptions and might appear to be merely a formal parameter of the geometric construction.

The geometric interpretation, however, does not deprive temperature of physical significance. A concrete question is how the temperature assigned to an isolated system changes immediately after a sudden modification of its Hamiltonian. The state is kept fixed, but the energy--coherence structure entering the thermodynamic construction changes. To be quantitative, consider an instantaneous perturbation
\begin{equation}
H(g)= H+gV
\end{equation}
at fixed state $\omega$, and denote by $\beta_\omega^{H(g)}$ the inverse temperature assigned to $\omega$ relative to the perturbed Hamiltonian.  Assuming that differentiation can be interchanged with the thermodynamic limit, its variation is
\begin{multline}\label{eq:response}
\left.\frac{d}{dg} \beta_\omega^{H(g)}\right|_{g=0}=\\
-\lim_{I\nearrow\mathbb Z}\frac{
	\omega_0(V^{(I)})-\tau(V^{(I)})+\beta_\omega^H\mathrm{Re}\bigl\langle H^{(I)},\bar V^{(I)-}_{(\omega)}(\beta_\omega^H)\bigr\rangle_{\omega_0}^{c}
}
{
	\|H^{(I)}\|_{\omega_0,{\rm inc}}^2
},
\end{multline}
where $\omega_0$ is the state at zero inverse temperature on the canonical orbit through $\omega$, and
\begin{equation}
\bar V^{(I)-}_{(\omega)}(\beta)=
\int_0^{1}d\lambda\, e^{\frac{\lambda}{2}\beta H^{(I)-}_{[\omega]}} V^{(I)-}_{(\omega)} e^{-\frac{\lambda}{2}\beta H^{(I)-}_{[\omega]}}.
\end{equation}
Here the notation $(\omega)$ emphasizes that the representative of a generic perturbation depends on the particular state at which the dressing is performed, whereas $H^{(I)-}_{[\omega]}$ is intrinsic to the leaf.

The response formula~\eqref{eq:response} quantifies the sensitivity of the thermodynamic coordinate to a deformation of the Hamiltonian while the physical state is kept fixed. Its first contribution, proportional to the density of $\omega_0(V^{(I)})-\tau(V^{(I)})$, captures an additive shift of the inverse temperature that is common to all states on the same canonical orbit and is caused by the redefinition of its infinite-temperature reference state. The second contribution accounts for the deformation of the canonical orbit connecting that state to $\omega$. If the perturbation preserves the leaf, its dressed representative commutes with the dressed Hamiltonian, so that $[V^{(I)}_{(\omega)},H^{(I)}_{(\omega)}]=0$ and $V^{(I)-}_{(\omega)}=V^{(I)-}_{[\omega]}$. It then follows that $\bar V^{(I)-}_{(\omega)}(\beta)=V^{(I)-}_{[\omega]}$, and the response formula reduces to
\begin{equation}\label{eq:response_reduced}
\begin{gathered}
\left.\tfrac{d}{dg}\beta_\omega^{H(g)}\right|_{g=0}=\tfrac{\tau(V)-\omega_0(V)}{\|H\|_{\omega_0,{\rm inc}}^2}-
\tfrac{(H,V)_{\omega_0}^{\rm inc}}{\|H\|_{\omega_0,{\rm inc}}^2}\, \beta_\omega^H \\
\text{(if }V \text{ preserves the leaf)}\, .
\end{gathered}
\end{equation}
Here the limit $I\nearrow\mathbb Z$ is understood, and we have suppressed the superscript $(I)$. Unless $V$ generates the canonical direction itself, such a perturbation would generally define an independent pseudolocal flow and hence make the leaf non-generic.

To make the content of the formula more transparent, it is instructive to consider the case in which, before the perturbation, $\omega$ is the Gibbs state of $H$ at inverse temperature $\beta$, so that $\beta_\omega^H=\beta$. In this case, $\omega_0$ is the tracial state $\tau$, and the response reduces to
\begin{equation}\label{eq:vareq}
\left.\frac{d\beta_\omega^{H(g)}}{dg}\right|_{g=0}=
-\frac{\left\langle H,V\right\rangle_{\tau}^{c}}{\left\langle H,H\right\rangle_{\tau}^{c}}\,\beta\, .
\end{equation}
This has the same structure as equation~\eqref{eq:response_reduced}, without, however, requiring the perturbation to preserve the leaf and hence without implying nongenericity. After quotienting by its null space, the thermodynamic-limit covariance density associated with the tracial state defines a Kubo--Mori--Bogoliubov inner product on the corresponding space of traceless pseudolocal generators\footnote{To the best of our understanding, the appearance of the tracial state in equation~\eqref{eq:vareq}, and more generally of $\omega_0$ in equation~\eqref{eq:response}, reflects the fact that the energy--coherence structure is a property of the entire leaf rather than of a particular state on it. At finite volume, in particular, the states of the optimal decomposition on the commuting leaf are orthogonal, and the energy--coherence structure does not distinguish among them. It is therefore natural, within this finite-volume picture, that the reference state entering the Kubo--Mori--Bogoliubov inner product is the tracial state, whose uniform populations place all these states on an equal footing. We expect the same qualitative logic to underlie equation~\eqref{eq:response}. A quantitative argument is, however, more subtle in the noncommuting case, because the states of the optimal decomposition are no longer orthogonal and are geometrically distinguished by their generally nonuniform pairwise overlaps. The response formula~\eqref{eq:response} suggests the relevance of an extension of the incoherent inner product for generic perturbations, but we leave its systematic development to future investigation.}. Thus Eq.~\eqref{eq:vareq} has the natural interpretation that, to first order, only the component of the perturbation parallel to the energy direction affects the assigned inverse temperature. For $V\propto H$, this is simply the expected compensation for a change in the overall energy scale. The representation in terms of inner products clearly indicates that the nonequilibrium inverse temperature is only sensitive to the component of a Hamiltonian perturbation that couples to the incoherent energy direction selected by the state. In this precise sense, the nonequilibrium inverse temperature probes the energy--incoherent structure of the state.

Equation~\eqref{eq:vareq} also provides a simple estimate of the nonequilibrium inverse temperature of an isolated system when its state is known to be close to equilibrium. Indeed, it implies 
\begin{equation}
\beta_\omega^{H} \approx \frac{\left\langle H,K_\omega \right\rangle_{\tau}^{c}}{\left\langle H,H\right\rangle_{\tau}^{c}} \, ,
\end{equation}
where $K_\omega$ is the modular Hamiltonian, which in finite volume is given by $K_\omega=-\log\rho_\omega$.

\section{Subsystems}\label{s:subsystems}

The flow formulation also clarifies the problem of subsystems. We start by describing our perspective in general terms, developing the details later in Section~\ref{ss:firstlaw}.  A subsystem $A$ is specified by its reduced density matrix $\rho_{\omega_A}$. The corresponding functional $\omega_A$ is the restriction of $\omega$ to the algebra $\mathcal A_A$ of local observables supported on the subsystem
\begin{equation}
\omega_A:O\mapsto \omega(O)\qquad O\in \mathcal A_A\, .
\end{equation}
In finite systems, the reduction of a minimum-variance leaf of the full system generically produces a high-dimensional subset of the subsystem state space\footnote{Let $\rho[\{p\}]=\sum_i p_i |\varphi_i\rangle\langle\varphi_i|$ be the family of density matrices on a leaf. The corresponding reduced density matrices on $A$ are $\rho_A[\{p\}]=\sum_i p_i\,\mathrm{tr}_{\bar A}(|\varphi_i\rangle\langle\varphi_i|)$. Thus the reduced image of a leaf is the convex hull of the reduced states $\mathrm{tr}_{\bar A}(|\varphi_i\rangle\langle\varphi_i|)$. Generically, its affine dimension is $\min(d-1,d_A^2-1)$, where $d$ is the Hilbert-space dimension of the full system and $d_A$ that of the subsystem. Hence, whenever $d_A^2\le d$, the reduced image is expected to be full-dimensional in the subsystem state space. For a spin-$\frac{1}{2}$ chain, this already happens when the subsystem contains no more than half of the chain.}.
In the thermodynamic-limit formulation, however, this full reduced set is not the relevant object. What survives locally is the reduced action of the pseudolocal flow. The global canonical direction therefore induces a local thermodynamic direction on the subsystem, obtained from \eqref{eq:canonical} by restricting the test observables $O$ to those supported in $A$, i.e., $O\in\mathcal A_A$
\begin{equation}\label{eq:reduction0}
\widehat {H_{A}^\omega}_{\omega_A}(O)\equiv \langle\!\langle H_A^\omega,O\rangle\!\rangle_{\omega_A}^{c}+\tfrac{i}{2}\omega_A([\tilde H^\omega_{A,[\omega_A]},O])=\widehat H_{\omega}(O).
\end{equation}
Here $\tilde H^\omega_{A,[\omega_A]}$ is the harmonic conjugate of $H_A^\omega$ with respect to the reduced state $\omega_A$. As we shall see in Section~\ref{ss:firstlaw}, $\tilde H^\omega_{A,[\omega_A]}$ and $H_A^\omega$ are fully determined by \eqref{eq:reduction0} ($H_A^\omega$ is determined up to an additive constant).

We envisage two ideal situations. The first is geometric: the reduced canonical direction is state-independent along a path ending at the infinite-temperature state of the subsystem, so that its inverse temperature can be defined as the coordinate of the reduced canonical flow. The second is dynamical: physical time evolution transports the effective subsystem leaf into compatible leaves, at least locally in time, equivalently preserving the reduced adjoint-action structure. In this case, the physical evolution could be decomposed into a unitary component, transverse to the effective leaf, and a temperature component, tangent to it. Neither situation, however, should be expected to hold exactly in general, and the obstruction is already visible when the global state is stationary. 

\subsection{Subsystems of systems in equilibrium}\label{ss:equi}

Even when the global state is at equilibrium, the subsystem Hamiltonian induced by the global energy flow depends explicitly on the temperature of the full state. This dependence is not accidental: the reduction to a subsystem cuts correlations between $A$ and its complement, and these correlations are themselves temperature-dependent.  Equivalently, the modular Hamiltonian of the reduced equilibrium state (linearly related to the ``Hamiltonian of mean force'' of Refs.~\cite{CampisiTalknerHanggi2009,TalknerHanggi2020,Trushechkin2022,Burke2024}) is not proportional to the bare subsystem Hamiltonian: for local interactions, the two Hamiltonians are expected to differ in a nontrivial way only near the boundaries of $A$, up to corrections controlled by the decay of correlations \cite{KlieschGogolinKastoryanoRieraEisert2014, CapelMoscolariTeufelWessel2025,HernandezSantana2015,HartmannGemmerMahlerHess2004}. From a purely thermodynamic perspective, this boundary sensitivity is not an obstruction. The inverse temperature is defined as the response of entropy to a variation of energy, and the subsystem is eventually taken to be asymptotically large. Boundary terms are then subextensive and can affect the inferred inverse temperature only through an uncertainty that scales as the boundary-to-volume ratio, hence as $1/|A|$ in the one-dimensional setting considered here. From a more foundational point of view, however, the same boundary terms are less innocuous. They prevent the temperature of a finite subsystem from being defined exactly and canonically. To the best of our knowledge, existing algebraic approaches do not provide a canonical assignment of an inverse temperature to a subsystem, even when the definition is allowed to depend on the embedding of the subsystem in the full system.

From our perspective, bulk and boundary contributions can be distinguished at the level of pseudolocal flows. To this end, we introduce a weaker notion of canonical flow.

\begin{description}

\item[Weak canonical flow]
We say that the canonical flow induces a weak canonical flow on an interval $A$ if, along the relevant segment of the canonical orbit, the induced local thermodynamic direction is independent of the point on the orbit up to boundary effects. More precisely, for observables of uniformly bounded support and unit norm, the discrepancy between the thermodynamic directions induced at different points of the orbit must decay summably with the distance of the support from the boundary of $A$. One possible way of making the notion precise is as follows. Let $\Delta_{\omega',\omega}^{A}=\widehat {H_A^{\omega'}}_{\omega_A}-\widehat{H_A^{\omega}}_{\omega_A}$, with $\omega'$ any state on the orbit, and define the local norm
\begin{equation}
\|\Phi\|_X=\sup\{|\Phi(O)|:O\in\mathcal A_X,\ \|O\|\le 1\}\, .
\end{equation}
Then, for every fixed $R$, there exists a summable function  $\epsilon_R(r)$, $\sum_{r\ge0}\epsilon_R(r)<\infty$, independent of $A$, $X$, and $\omega'$, such that
\begin{equation}
\|\Delta_{\omega',\omega}^{A}\|_X\le\,\epsilon_R(r)
\end{equation}
for all $X\subset A$ with $|X|\le R$, where $r=\mathrm{dist}(X,\partial A)$.

\end{description}

We propose this weaker definition as a way of assigning a temperature to a subsystem of a system in thermal equilibrium\footnote{This construction is reminiscent in spirit of the framework based on thermal observables of Ref.~\cite{Buchholz2002Thermodynamic}, although the role of the thermal observables is here played by the bulk action of the weak canonical flow.}. Indeed, along the finite segment of the canonical orbit connecting a thermal state to the infinite-temperature state, the correlation length remains uniformly bounded. Thus the canonical flow through an equilibrium state induces a weak canonical flow on the subsystem. This leads to the identification, up to boundary ambiguities, of the subsystem temperature with the system one. This can be seen as an alternative way of expressing the statement  that temperature is a local property~\cite{KlieschGogolinKastoryanoRieraEisert2014}.

We stress, however, that the weak canonical flow does not determine an exact leaf: it disregards residual curvature arising from the $\beta$-dependence of the effective subsystem Hamiltonian near the boundary of $A$. The resulting $O(|A|^{-1})$ ambiguity is structural, rather than metrological. Unlike the limitations on temperature-measurement precision discussed in Ref.~\cite{DePasquale2016Local}, it is tied to the reduction to a subsystem. In the rest of the paper, this ambiguity will be taken for granted.

As shown in the next subsection, this provides the required anchor for defining a subsystem temperature out of equilibrium. 

\subsection{Canonical flow induced on subsystems}\label{ss:firstlaw}

\begin{figure}
\centering
\includegraphics[width=0.35
\textwidth]{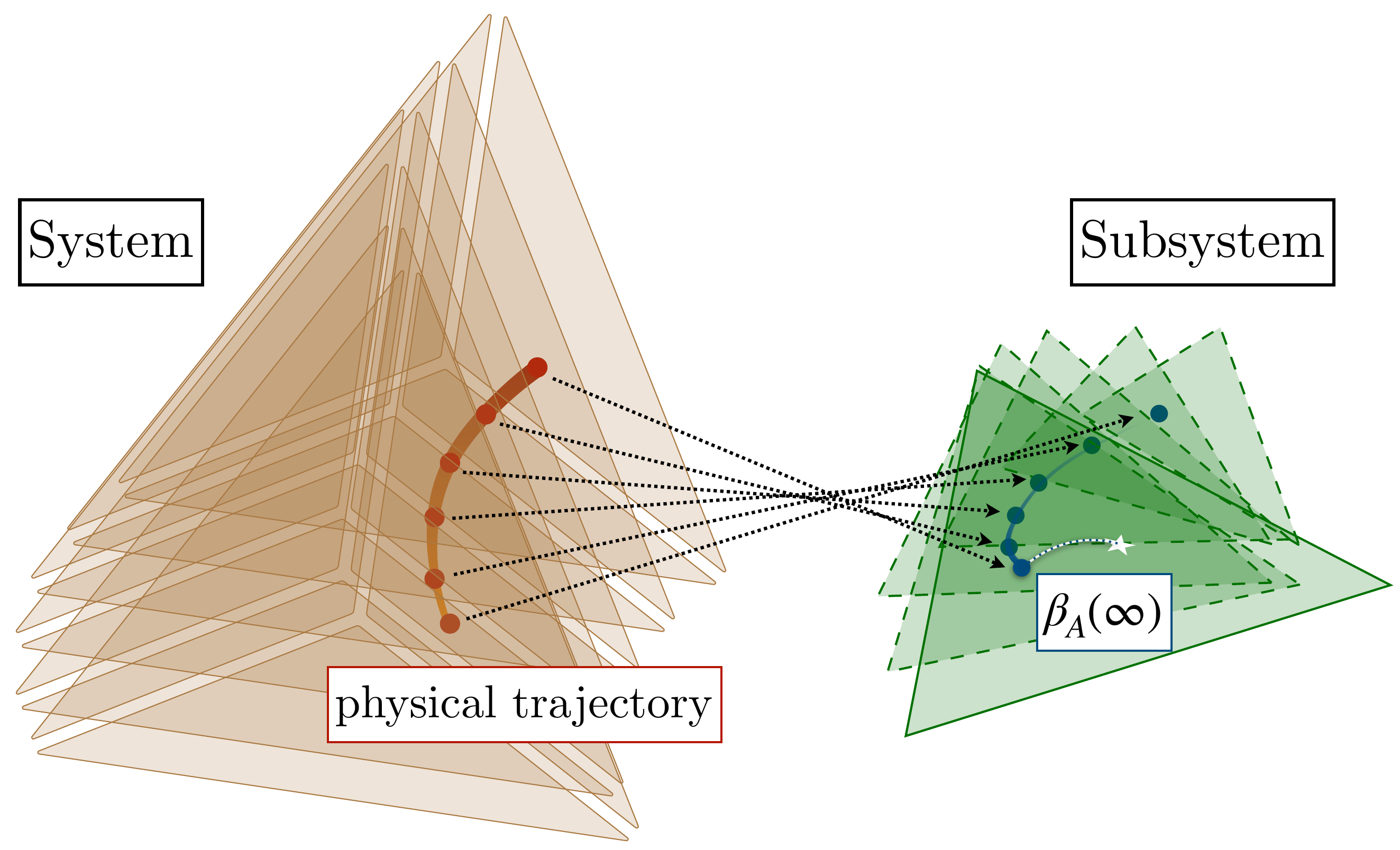}
\hspace{1cm}
\includegraphics[width=0.3
\textwidth]{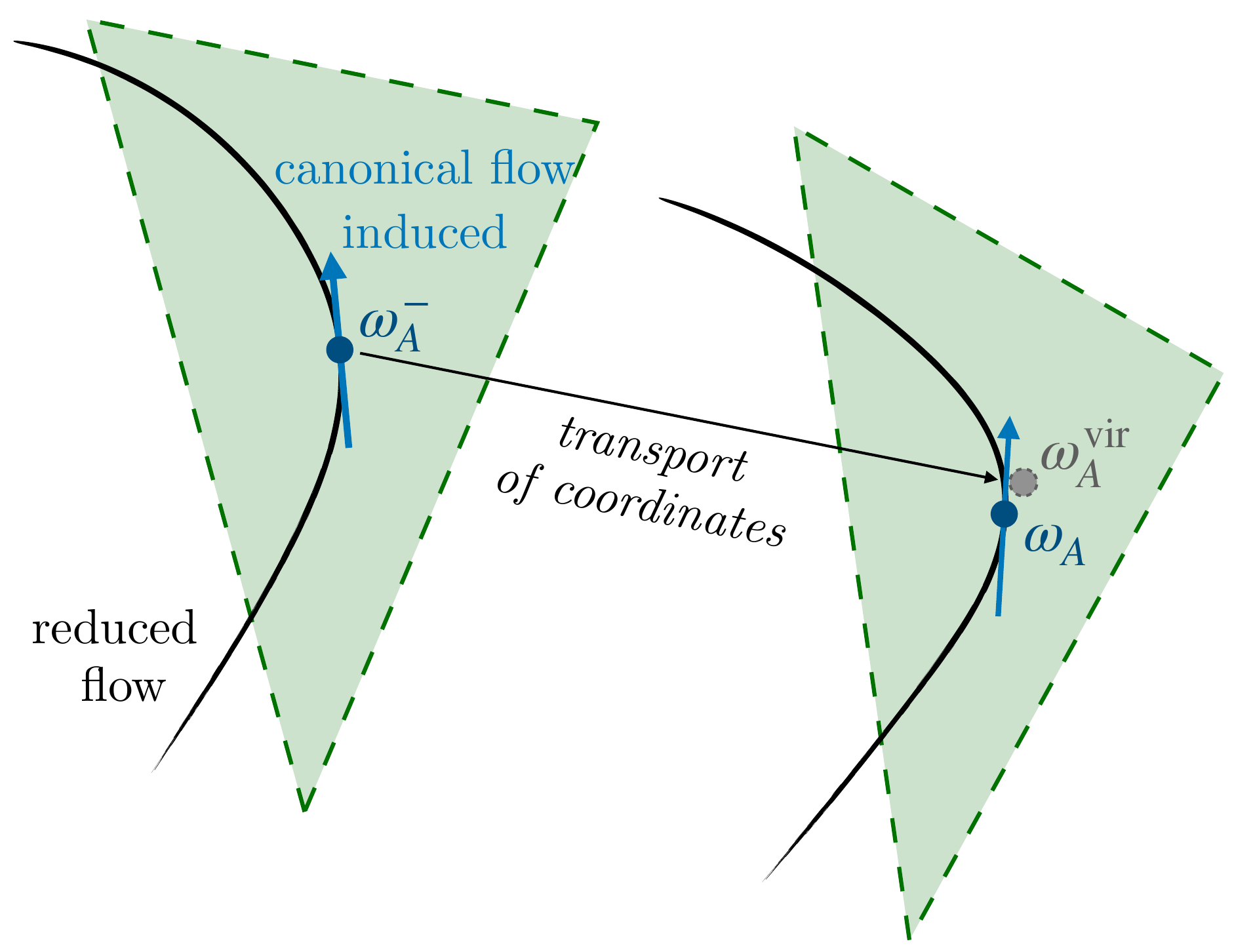}
\caption{\textbf{Geometric construction of subsystem temperature.} 
\textsc{Top}: The global state evolves through the leaves of the full-system foliation, represented as (brown) tetrahedra as if the state space were four-dimensional. Reduction maps this trajectory to the subsystem state space, represented as three-dimensional, where the effective leaves are shown as (green) triangles. The global canonical flow induces a local thermodynamic direction on the subsystem, fixing the time derivative of the subsystem inverse temperature. 
The actual value of the inverse temperature is then obtained by integrating its rate from an admissible reference state at which the induced canonical flow is weak, which in the cartoon is assumed to be the reduced infinite-time state.
\textsc{Bottom}: The subsystem state space is represented in three dimensions. The orbit of the canonical flow through the global state defines, at each time, an effective leaf and an induced canonical direction in the subsystem state space. After an infinitesimal time $dt$, the effective leaf changes. The state at time $t-dt$ is transported to a virtual state on the new leaf by keeping the same leaf coordinates. Comparing this virtual state with the actual state at time $t$ along the induced canonical flow determines the infinitesimal change of the subsystem inverse temperature.
}\label{f:submap}
\end{figure}

It is useful to return momentarily to finite systems. For an isolated system with nondegenerate energies, the population coordinates $\{p_i\}$, together with the ordering of the corresponding states by the expectation value of the Hamiltonian, define a natural notion of parallel transport of tangent directions between leaves of the foliation. For a subsystem, an analogous construction remains possible only locally in time, because the effective foliation itself changes along the physical trajectory and its leaves are not, in general, mutually compatible. To simplify the notations, we will indicate with $\omega$  the state at time $t$ and with $\omega^{-}$ the state at  the infinitesimally earlier time $t-dt$.

The procedure we propose consists of three steps:

\begin{enumerate}
\item At time $t$, we construct the effective leaf $\mathcal L_{H_A^{\omega}}(\omega_A)$ through the subsystem state $\omega_{A}$.
\item On the effective leaf at time $t$, we introduce a virtual state $\omega_{A}^{\rm vir}$ corresponding to the actual subsystem state at the infinitesimally earlier time $t-dt$. In finite volume, this correspondence is defined by parallel transport of the leaf population coordinates. Since the thermodynamic entropy depends only on the populations, we have
\begin{equation}\label{eq:virtual}
S_{\rm th}(\omega_{A}^{\rm vir}|\mathcal L_{H_A^{\omega}}(\omega_A))=
S_{\rm th}(\omega_A^{-}|\mathcal L_{H_A^{\omega^{-}}}(\omega_A^{-}))\, .
\end{equation}
\item As $dt\to0$, the separation between the actual state $\omega_{A}$ and the virtual state $\omega_{A}^{\rm vir}$ selects a direction on the effective leaf which, at finite volume, generally also has a component transverse to the canonical one. For a sufficiently large subsystem, we make what we shall call the \emph{alignment assumption}: the contribution  from the transverse component to the variation of the thermodynamic entropy is subleading. A breakdown of this assumption would undermine the assignment of a single inverse-temperature coordinate to the subsystem. In the thermodynamic limit, where the individual populations have no counterpart, we therefore identify $\omega_{A}^{\rm vir}$ with the point on the regular canonical orbit satisfying \eqref{eq:virtual}. Its separation from $\omega_{A}$ is then a displacement $\delta\beta_A$ of the inverse-temperature coordinate, determined by \eqref{eq:virtual} through the susceptibility $\partial_{\beta_A} S_{\rm th}$; see Figure~\ref{f:submap}.

\end{enumerate}

In finite volume, the first step is elementary, and the solution reads 
\begin{equation}\label{eq:HAHtildeA}
\begin{aligned}
&H_{A}^\omega-\omega_A(H_{A}^\omega)\mathrm I=\frac{1}{2}\{\rho_{\omega_A}^{-1},D^H(\omega;A)\}\\
&\tilde H^\omega_{A,[\omega_A]}=-\frac{i}{2}[\rho_{\omega_A}^{-1},D^H(\omega;A)]\, ,
\end{aligned}
\end{equation}
where
\begin{equation}\label{eq:DH}
D^H(\omega;A)=\frac{1}{2} \mathrm{tr}_{\bar A}\bigl(\{H-\omega(H),\rho_\omega\}-i[\tilde H_{[\omega]},\rho_\omega]\bigr)\, .
\end{equation}

The second and the third steps are discussed in Appendix~\ref{a:subsystem}. In conclusion, the alignment assumption results in
\begin{equation}\label{eq:betadot}
\frac{d\beta_A}{dt}=\frac{\frac{d}{dt}S_{\rm th}(\omega_A|\mathcal L_{H_A^\omega}(\omega_A))}{\partial_{\beta_A}S_{\rm th}(\omega_A|\mathcal L_{H_A^\omega}(\omega_A))}\, .
\end{equation}

Before discussing the meaning of the two terms in this equation, we state the scaling expectation underlying the macroscopic-time interpretation of this equation. For an interval $A$, we expect $dS_{\mathrm{th}}/dt$ to be a boundary contribution of order $O(1)$, whereas the susceptibility in the denominator should be of order $O(|A|)$ away from singular points. Under these assumptions, $d\beta_A/dt=O(|A|^{-1})$. Thus the subsystem inverse temperature evolves on a macroscopic timescale that grows at least linearly with the subsystem size. 

\subsection{Flow of local inverse temperature}\label{ss:lowsub}

The aim of this section is to simplify the dynamical equation for the subsystem inverse temperature~\eqref{eq:betadot} by evaluating the partial derivative and expressing everything in terms of a single quantity: the leaf modular Hamiltonian associated with the effective subsystem leaf.  The starting point is the $\omega_A$-dressed Hamiltonian $H^\omega_{A,(\omega_A)}$, which can be written as $H^\omega_{A,(\omega_A)}=\rho_{\omega_A}^{-1/2}D^H(\omega;A)\rho_{\omega_A}^{-1/2}$, where $D^H(\omega;A)$ is defined in \eqref{eq:DH}. Let $|\Psi_i^{A,\omega}\rangle$ be the eigenvectors of $H^\omega_{A,(\omega_A)}$ and define
\begin{equation}
p_i^{A,\omega}=\langle\Psi_i^{A,\omega}|\rho_{\omega_A}|\Psi_i^{A,\omega}\rangle.
\end{equation}
We identify the $\omega_A$-dressed leaf modular Hamiltonian with the modular Hamiltonian of the diagonal ensemble associated with $\omega_A$ in the basis $\{|\Psi_i^{A,\omega}\rangle\}$. 
\begin{equation}
K^{\mathcal L,\omega}_{A,(\omega_A)}=-\sum_i \log p_i^{A,\omega}\,|\Psi_i^{A,\omega}\rangle\langle\Psi_i^{A,\omega}|.
\end{equation}
Note that $K^{\mathcal L,\omega}_{A,(\omega_A)}$ can be interpreted as the entanglement Hamiltonian of the infinite time-averaged state under the dynamics generated by $H^\omega_{A,(\omega_A)}$.

We then define the corresponding subsystem leaf modular Hamiltonian $K_A^{{\mathcal L},\omega}$ as the operator whose $\omega_A$-dressed representative is $K^{\mathcal L,\omega}_{A,(\omega_A)}$. Thus $K_A^{\mathcal L,\omega}$ is determined by the Lyapunov equation
\begin{equation}
\frac{1}{2}\bigl\{\rho_{\omega_A},K^{\mathcal L,\omega}_{A,(\omega_A)}\bigr\}=
\rho_{\omega_A}^{\frac{1}{2}}K_A^{\mathcal L,\omega}\rho_{\omega_A}^{\frac{1}{2}}.
\end{equation}
Equivalently,
\begin{equation}
K_A^{\mathcal L,\omega}=
\frac{1}{2}
\bigl(
	\rho_{\omega_A}^{-\frac{1}{2}}K^{\mathcal L,\omega}_{A,(\omega_A)}\rho_{\omega_A}^{\frac{1}{2}}+
	\rho_{\omega_A}^{\frac{1}{2}}K^{\mathcal L,\omega}_{A,(\omega_A)}\rho_{\omega_A}^{-\frac{1}{2}}
\bigr)\, .
\end{equation}
This operator plays, on a non-commuting effective leaf, the role played by the modular Hamiltonian on the commuting leaf.

The thermodynamic entropy is then
\begin{multline}
S_{\rm th}(\omega_A|\mathcal L_{H_A^\omega}(\omega_A))=-\sum_i p_i^{A,\omega}\log p_i^{A,\omega}\\
=\omega_A\bigl(K^{\mathcal L,\omega}_{A,(\omega_A)}\bigr)=\omega_A\bigl(K_A^{\mathcal L,\omega}\bigr).
\end{multline}
Its derivative along the reduced canonical direction can be written as the canonical flow of the subsystem leaf modular Hamiltonian:
\begin{multline}
\partial_{\beta_A} S_{\rm th}(\omega_A|\mathcal L_{H_A^\omega}(\omega_A))=\\
-\widehat {H_{A}^\omega}_{\omega_A}(K_A^{\mathcal L,\omega})=
-\widehat H_{\omega}(K_A^{\mathcal L,\omega}\otimes \mathbf 1_{\bar A}).
\end{multline}
The last identity follows from the definition of induced canonical flow~\eqref{eq:reduction} (note that on the right-hand-side of the equation the operator $K_A^{\mathcal L,\omega}$ has been extended trivially to the complement of $A$). Therefore \eqref{eq:betadot} can be written as
\begin{equation}\label{eq:firstlaw}
\frac{d}{dt} S_{\rm th}(\omega_A|\mathcal L_{H_A^\omega}(\omega_A))+\widehat {H_{A}^\omega}_{\omega_A}(K_A^{\mathcal L,\omega})\frac{d \beta_A}{dt}=0 
\end{equation}
or, alternatively,  as an equation for the subsystem leaf modular Hamiltonian
\begin{equation}\label{eq:firstlaw1}
\frac{d}{dt}  [\omega\!\left(K_A^{\mathcal L,\omega}\otimes \mathbf 1_{\bar A}\right)]+\widehat H_{\omega}\!\left(K_A^{\mathcal L,\omega}\otimes \mathbf 1_{\bar A}\right)\frac{d \beta_A}{dt}=0\, .
\end{equation}
We stress that the inverse temperature $\beta_A$ appears in this equation only explicitly through its time derivative. In addition, all the other quantities in the equation depend on the state locally in time and space (only the behaviour of the state in a large-enough subsystem surrounding $A$ is relevant), hence we can regard $\frac{d \beta_A}{dt}$ as a local property of the state. 

\paragraph{The reference state}

We cannot draw the same conclusion for the inverse temperature $\beta_A$, which, to be fixed, requires a reference state as an anchor. 
To simplify the notations, the time of the reference state will be called $t_0$ and the state $\omega^{(0)}$, i.e., $\omega^{(0)}=\omega\circ\sigma^{t_0-t}$.
We identify two kinds of reference states. First, if the subsystem $A$ is much larger than 

\begin{itemize}

\item[-] the typical range of the local densities, with support intersecting $A$, of the harmonic conjugate $\tilde H_{[\omega^{(0)}]}$ of $H$

\item[-] the correlation length of the states on the orbit of the canonical flow joining $\omega^{(0)}$ with the corresponding infinite temperature state,

\end{itemize}

\noindent then the state at $t_0$ provides an anchor: the canonical flow induces a weak canonical flow on $A$ (cf. Section~\ref{ss:equi}) and, in turn, $\beta_A(t_0)$ can be fixed with an indetermination $O(1/|A|)$. Second, if the assumption of local relaxation is met, the state approached at infinite time belongs to the orbit of  a weak canonical flow. We stress that, in contrast with the equilibrium case, these anchors do not lead to the identification of the subsystem temperature with the system one. The subtlety is in the characterization of the infinite-temperature state. Indeed, while the tracial state $\tau$ is an equilibrium state that corresponds to the infinite-temperature state also in the subsystem, out of equilibrium there is no reason to expect $(\omega^{(0)})_{\beta=0}(H_A^{\omega^{(0)}}\otimes \mathbf 1_{\bar A})=\tau(H_A^{\omega^{(0)}}\otimes \mathbf 1_{\bar A})$. There will be instead an inverse temperature $\beta_{\tau,A}(t_0)$, generically different from $0$, such that 
\begin{equation}
(\omega^{(0)})_{\beta_{\tau, A}(t_0)}(H^{\omega^{(0)}}_A\otimes \mathbf 1_{\bar A})=\tau(H_A^{\omega^{(0)}}\otimes \mathbf 1_{\bar A})\, .
\end{equation}
At the anchor, the subsystem inverse temperature $\beta_A(t_0)$ (with $t_0$ possibly equal to infinity) will therefore be given by
 \begin{equation}
 \beta_A(t_0)=\beta_{\omega^{(0)}}^H-\beta_{\tau, A}(t_0)=\beta_\omega^H-\beta_{\tau, A}(t_0)\, .
 \end{equation}
This is consistent with the simple expectation that subsystems of a nonequilibrium state  can be at different temperatures.  We do not prove here that integrating the local rate from different admissible anchors gives the same inverse temperature. We expect any such discrepancy to be of the same order as the intrinsic boundary ambiguity, namely $O(|A|^{-1})$ in one dimension; establishing this anchor consistency is however an open problem.

\section{Examples}\label{s:examples}

In this section we present a few examples intended to clarify the physical content of the framework and to provide a practical outlook on aspects that have been only marginally explored here and deserve separate investigation. Our aim is to balance their illustrative value with their precise theoretical status. Accordingly, we do not attempt to verify every assumption entering the general construction when the point at issue can be illustrated more transparently in a simplified setting. In particular, we do not verify that the parameters considered lie within the regularity region discussed in Appendix~\ref{a:leafthrough}, both because we have not derived explicit bounds and because we expect such bounds to provide sufficient conditions for regularity rather than physically meaningful thresholds.

The first simplification is in the choice of the initial state. We set it to be a thermal state of either one of the two trivial Hamiltonians
\begin{equation}\label{eq:H0pf}
H_{0,p}^{(I)}=-\frac{J_0}{2} \sum_{\ell\in I} \sigma_\ell^z\qquad H_{0,f}^{(I)}=-\frac{J_0}{2} \sum_{\ell\in I} \sigma_\ell^z\sigma_{\ell+1}^z\, .
\end{equation}
This choice has the advantage of making transparent both the existence of the thermodynamic limit and the regularity properties that, in the general discussion, were implicitly justified using the $C^\star$-algebraic formulation of local observables.  At zero temperature (we assume $J_0>0$) $H_{0,p}^{(I)}$ describes a paramagnetic phase, whereas $H_{0,f}^{(I)}$ describes a ferromagnetic one. 

\subsection{Generic and non-generic leaves with an integrable Hamiltonian}

\begin{figure*}
\centering
\includegraphics[width=0.85\textwidth]{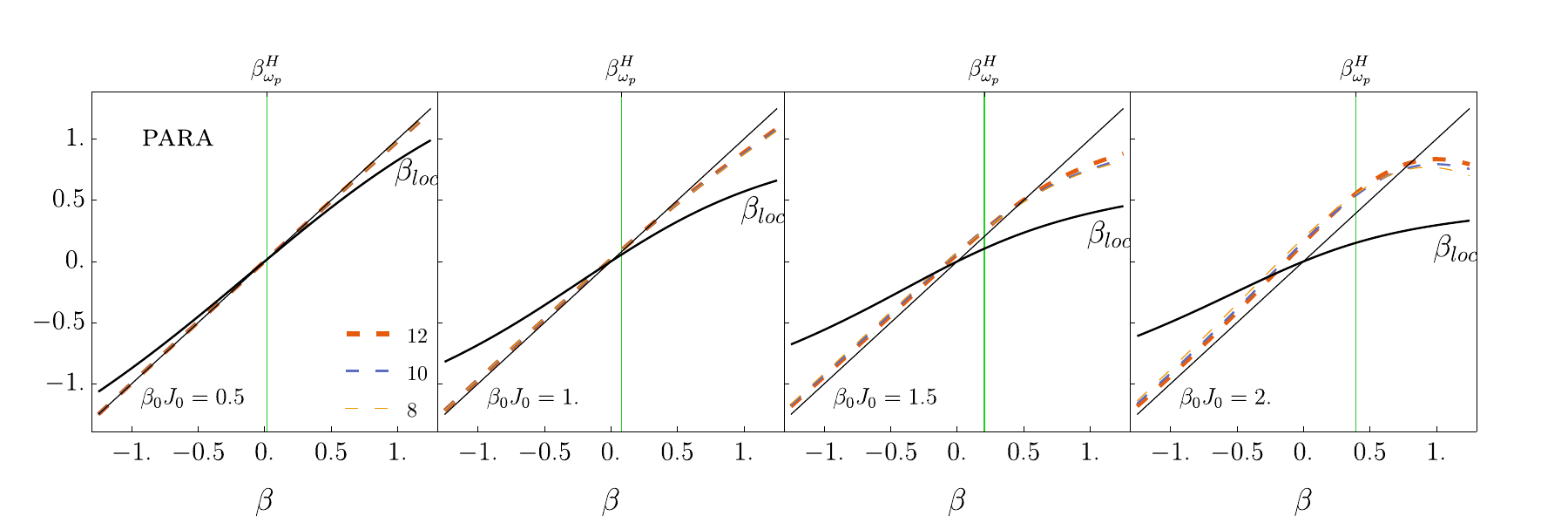}
\caption{\textbf{Temperature vs local temperature vs bare temperature.} 
The stabilized late-time local inverse temperature $\beta_{\mathrm{loc}}$ (black solid curves) as a function of the nonequilibrium inverse temperature $\beta$ in the leaves across  thermal states (with inverse temperatures $\beta_0$) of the paramagnetic pre-quench Hamiltonian $H_{0,p}^{(I)}$ \eqref{eq:H0pf}. The coupling constants of the Hamiltonian \eqref{eq:exampleH} are set to $J=1$, $\gamma=-2$, $h=0.5$, and $D=0.75$. The dashed curves are the estimators $\beta^{(I)}_{\mathrm{bare}}(\beta)$ (bare inverse temperatures), defined in \eqref{eq:dressing}, for three spin chains of lengths $|I|=8,10,12$ and periodic boundary conditions. The thin straight black line is just shown for reference, as it represents $\beta\mapsto \beta$.
}\label{f:example1}
\end{figure*}

For the sake of simplicity, we consider a noninteracting Hamiltonian. We choose a rotated XY model in a transverse field and with a Dzyaloshinskii--Moriya interaction
\begin{multline}\label{eq:exampleH}
H^{(I)}=\frac{J}{2} \sum_{\ell\in I} \frac{1+\gamma}{2} \sigma_\ell^z\sigma_{\ell+1}^z+\frac{1-\gamma}{2}  \sigma_\ell^x\sigma_{\ell+1}^x\\
+ h \sigma_\ell^y+D( \sigma_\ell^z\sigma_{\ell+1}^x- \sigma_\ell^x\sigma_{\ell+1}^z)\, .
\end{multline}
The (density of the) $\omega$-dressed Hamiltonians can be readily computed using the representation \eqref{eq:hw-def}. We find 
\begin{equation}
\begin{aligned}
h_{(\omega_p),\ell}=&
J\Bigl[\gamma_x\sigma_\ell^x\sigma_{\ell+1}^x+\gamma_y\sigma_\ell^y\sigma_{\ell+1}^y+\tfrac{1+\gamma}{4}\sigma_\ell^z\sigma_{\ell+1}^z+h_y^p \sigma_\ell^y\\
&\quad+D_y(\sigma_\ell^z\sigma_{\ell+1}^x-\sigma_\ell^x\sigma_{\ell+1}^z)\Bigr]\\
h_{(\omega_f),\ell}=&
J\Bigl[\gamma_x\sigma_\ell^x\sigma_{\ell+1}^x+\gamma_{y}\sigma_{\ell-1}^z\sigma_\ell^y\sigma_{\ell+1}^y\sigma_{\ell+2}^z+\tfrac{1+\gamma}{4}\sigma_\ell^z\sigma_{\ell+1}^z\\
&\quad+h_y^f \sigma_\ell^y\tfrac{\mathbf 1+\sigma_{\ell-1}^z\sigma_{\ell+1}^z}{2}+\tfrac{D}{2}(\sigma_\ell^z\sigma_{\ell+1}^x-\sigma_\ell^x\sigma_{\ell+1}^z)\\
&\quad+\tfrac{h}{2}\sigma_\ell^y\tfrac{\mathbf 1-\sigma_{\ell-1}^z\sigma_{\ell+1}^z}{2}\Bigr]
\end{aligned}
\end{equation}
where the coefficients read
\begin{equation}
\begin{gathered}
\gamma_x=\tfrac{1-\gamma}{4}\tfrac{\cosh(\beta_0 J_0)+1}{2\cosh(\beta_0 J_0)}\quad
\gamma_y=\tfrac{1-\gamma}{4}\tfrac{\cosh(\beta_0 J_0)-1}{2\cosh(\beta_0 J_0)}\\ 
h_y^p=\tfrac{h}{2\cosh(\frac{\beta_0 J_0}{2})}\quad
h_y^f=\tfrac{h}{2\cosh(\beta_0 J_0)}\quad
D_y= \tfrac{D}{2\cosh(\frac{\beta_0 J_0}{2})}\, ,
\end{gathered}
\end{equation}
and $\omega_{p}$ and $\omega_{f}$  denote the thermal states associated with  $H_{0,p}^{(I)}$ and $H_{0,f}^{(I)}$---\eqref{eq:H0pf}, respectively. 

Except for a $\mathbb Z_2$ symmetry of spin flip in the $y$ direction, for a generic choice of the parameters, $H_{(\omega_p)}$  is expected to possess no independent local or quasilocal conserved quantities beyond the energy. Note, however, that there are parameters that make the leaf non-generic; for example,  it is integrable for $h=D=0$ (XYZ Heisenberg model) or for $\gamma=1$ (transverse-field Ising model with a Dzyaloshinskii--Moriya interaction). On the other hand, $H_{(\omega_f)}$ is always noninteracting since both the pre-quench and the post-quench Hamiltonians are quadratic in the same Jordan-Wigner fermions---cf.~\eqref{eq:hw-def}.   Thus, for the same integrable post-quench Hamiltonian, different simple initial thermal states can be associated with generic and non-generic leaves.

We also report the density of the harmonic conjugates $\tilde H_{[\omega_p]}$ and  $\tilde H_{[\omega_f]}$
\begin{equation}
\begin{aligned}
\tilde h_{(\omega_p),\ell}=&-J\eta\Bigl[\tfrac{1}{4}\tfrac{1-\gamma}{1+\eta^2}(\sigma_\ell^x\sigma_{\ell+1}^y+\sigma_\ell^y\sigma_{\ell+1}^x)
-\tfrac{h}{2}\sigma_\ell^x\\
&\quad+\tfrac{D}{2}(\sigma_\ell^z\sigma_{\ell+1}^y-\sigma_\ell^y\sigma_{\ell+1}^z)\Bigr]\\
\tilde h_{(\omega_f),\ell}=&-J\eta\Bigl[\tfrac{1}{4}\tfrac{1-\gamma}{1+\eta^2}(\sigma_\ell^x\sigma_{\ell+1}^y\sigma_{\ell+2}^z+\sigma_{\ell-1}^z\sigma_\ell^y\sigma_{\ell+1}^x)\\
&\quad -\tfrac{1}{2}\tfrac{h}{1+\eta^2}(\sigma_\ell^z\sigma_{\ell+1}^x+\sigma_\ell^x\sigma_{\ell+1}^z)\Bigr]\, ,
\end{aligned}
\end{equation}
where $\eta=\tanh(\frac{\beta_0J_0}{2})$. Note that the densities of the harmonic conjugates are local in the strict sense because  the pre-quench Hamiltonians are sums of mutually commuting terms with finite range.

\paragraph{Stabilized late-time local inverse temperature.}

For generic parameters,  the inverse temperature can be computed by solving \eqref{eq:condT0} in finite chains and extrapolating the solution to the thermodynamic limit. In practice, the convergence is so rapid that finite-size effects are already negligible at sizes accessible to exact diagonalization. Translational invariance further reduces \eqref{eq:condT0} to an equation involving the energy density. Nevertheless, because the Hamiltonian is integrable, local observables relax to stationary values described by a generalized Gibbs ensemble rather than by a canonical ensemble. The coefficient conjugate to the Hamiltonian in such an ensemble does not, by itself, define an invariant temperature: independent linear recombinations of the conserved charges induce corresponding linear transformations of their conjugate coordinates. We resolve this ambiguity by defining the inverse temperature through stability under infinitesimal integrability breaking. More precisely, we first add a generic perturbation of arbitrarily small strength, take the infinite-time limit at fixed perturbation, and only then let the perturbation strength vanish. Assuming continuity of the resulting thermal state, the stabilized inverse temperature $\beta_{\mathrm{loc}}$ is therefore the inverse temperature of the Gibbs ensemble having the same energy density as $\omega$. This definition is stable under infinitesimal generic perturbations of the Hamiltonian. We stress that this prescription does not conflict with the need for a generalized Gibbs ensemble to describe late-time local observables in the integrable model. The remaining conserved charges and their conjugate coordinates should instead be chosen so as to complement the distinguished energy coordinate fixed by $\beta_{\mathrm{loc}}$. We are then in a position to compare the nonequilibrium inverse temperature $\beta $ (at fixed energy coherence), which is independent of time by definition,  with the stabilized late-time local inverse temperature $\beta_{\mathrm{loc}}$---Figure~\ref{f:example1}. The figure shows  that the stabilized late-time local temperature is larger than the nonequilibrium temperature in absolute value (note that the plots show the inverse of the temperatures). This is intuitive, as the stabilized late-time local temperature carries also a contribution from the loss of information about the energy coherence  of the state.  However, we have not investigated whether such inequality is always satisfied (we are comparing different leaves). 

\paragraph{Bias of the maximum-entropy estimator.}

\begin{table}[t]
\begin{center}
\begin{tabular}{c|c|c}
$|I|$ & $\beta_\omega^H$ & $\beta_{\mathrm{bare}}^{(I)}$\\
\hline
6  & 0.39462 & 0.526\\
7  & 0.39468 & 0.493\\
8  & 0.39469 & 0.552\\
9  & 0.39469 & 0.564\\
10 & 0.39469 & 0.540\\
11 & 0.39469 & 0.541\\
12 & 0.39469 & 0.555\\
13 & 0.39469 & 0.533\\
14 & 0.39469 & 0.532
\end{tabular}
\caption{Nonequilibrium inverse temperature $\beta_\omega^H$ and  bare inverse temperature $\beta_{\mathrm{bare}}^{(I)}$ for several chain lengths, in the thermal state of $H_{0,p}^{(I)}$ \eqref{eq:H0pf} at $\beta_0J_0=2$. The post-quench Hamiltonian \eqref{eq:exampleH} has the same couplings as in Fig.~\ref{f:example1}, with periodic boundary conditions.}\label{t:1}
\end{center}
\end{table}
Figure~\ref{f:example1} also shows the bare inverse temperature~\eqref{eq:dressing} as a function of the inverse temperature. The two quantities remain close when $|\beta_\omega^H|$ is small, but a systematic bias of the bare estimator becomes visible as $|\beta_\omega^H|$ increases. For a more direct comparison, Table~\ref{t:1} reports finite-chain estimates of the nonequilibrium inverse temperature $\beta_\omega^H$ and of the bare inverse temperature $\beta_{\mathrm{bare}}^{(I)}$ for the paramagnetic initial state with $\beta_0 J_0=2$ and periodic boundary conditions. 
The estimate of $\beta_\omega^H$ converges so rapidly that it is already practically indistinguishable from its thermodynamic-limit value in very short chains. The bare estimator exhibits larger and nonmonotonic finite-size effects. Extrapolations we have performed indicate a nonvanishing gap, but we do not report them, since we have no justification for any particular functional form. The raw data already show that the separation from $\beta_\omega^H$ is substantially larger than the observed finite-size fluctuations and persists at every accessible size, consistently with the bias visible in Fig.~\ref{f:example1}. A finite-size origin of the bias would therefore require a qualitative crossover beyond the accessible size range. We thus interpret the data as numerical evidence that the maximum-entropy prescription of Ref.~\cite{Fagotti2026Quantum} does not, on a generic noncommuting leaf, identify the inverse temperature associated with the canonical flow through $\omega$, as discussed in Section~\ref{s:isolated}. 

\paragraph{Energy coherence.}

Concerning the energy coherence content of the leaf, a strictly local operator encodes the density of quantum Fisher information of the paramagnetic state, or of the ferromagnetic state, with respect to the Hamiltonian, $\phi^{p/f}_{\ell}=2i [H,\tilde h_{(\omega_{p/f}),\ell}]$ (cf.~\eqref{eq:QFI}). 
We remark that the quantum Fisher information per unit length is equal to $\omega(\phi^{p/f}_\ell)$ in any state of the leaf. Since $\phi^{p/f}_{\ell}$ is a bounded operator, this immediately implies that the density of the QFI is finite even in the limits $\beta\rightarrow\pm\infty$, and, in finite volume, that the energy variance of every state of the optimal decomposition is at most proportional to the volume. This is a general feature: all the states in a leaf across a regular state have regular QFI with respect to the energy, that is to say, the energy QFI is at most extensive. Note that the corresponding operator at time $t$ is given by $\phi^{p/f}_{\ell}(t)=\sigma^{-t}(\phi^{p/f}_{\ell})$, resulting in a stationary QFI.

\subsection{Non-generic leaves with nonintegrable Hamiltonian}

Let us consider again the Hamiltonian $H^{(I)}$ with the additional perturbation
\begin{equation}\label{eq:exampleV}
V^{(I)}=\frac{J\delta_y}{2}\sum_{\ell\in I} \sigma_\ell^y\sigma_{\ell+1}^y\, .
\end{equation}
We focus on the paramagnetic state. The coupling constants of the $\omega$-dressed Hamiltonian change as follows
\begin{equation}
\begin{aligned}
\gamma_y\rightarrow &\gamma_y+\delta_y\tfrac{\cosh(\beta_0 J_0)+1}{4\cosh(\beta_0 J_0)}\\
\gamma_x\rightarrow &\gamma_x+\delta_y\tfrac{\cosh(\beta_0 J_0)-1}{4\cosh(\beta_0 J_0)}\, .
\end{aligned}
\end{equation}
If $\delta_y$ is given by
\begin{equation}
\delta_y=(1+\gamma)\tfrac{\cosh(\beta_0 J_0)}{\cosh(\beta_0 J_0)-1}-\tfrac{1-\gamma}{2}\tfrac{\cosh(\beta_0 J_0)+1}{\cosh(\beta_0 J_0)-1}
\end{equation}
then the $\omega$-dressed Hamiltonian is integrable (an XXZ Heisenberg model in a longitudinal field, up to a Dzyaloshinskii--Moriya term along the symmetry axis that can be removed by a site-dependent rotation~\cite{Alcaraz1990The}). On the other hand, the Hamiltonian reads
\begin{multline}
\tfrac{J}{2} \sum_{\ell\in I} \tfrac{1+\gamma}{2} \sigma_\ell^z\sigma_{\ell+1}^z+\tfrac{1-\gamma}{2}  \sigma_\ell^x\sigma_{\ell+1}^x+D( \sigma_\ell^z\sigma_{\ell+1}^x- \sigma_\ell^x\sigma_{\ell+1}^z)\\
+\Bigl((1+\gamma)\tfrac{\cosh(\beta_0 J_0)}{\cosh(\beta_0 J_0)-1}-\tfrac{1-\gamma}{2}\tfrac{\cosh(\beta_0 J_0)+1}{\cosh(\beta_0 J_0)-1}\Bigr)\sigma_\ell^y\sigma_{\ell+1}^y+ h \sigma_\ell^y
\end{multline}
which is instead expected to be nonintegrable for generic choices of the parameters (an XYZ-type model with Dzyaloshinskii--Moriya interaction and an external field). This explicitly shows that a nonintegrable post-quench Hamiltonian can support a non-generic, in this case integrable, leaf.

\subsection{Bipartitioning protocol}
 
 \begin{figure*}
\centering
\includegraphics[width=0.7\textwidth]{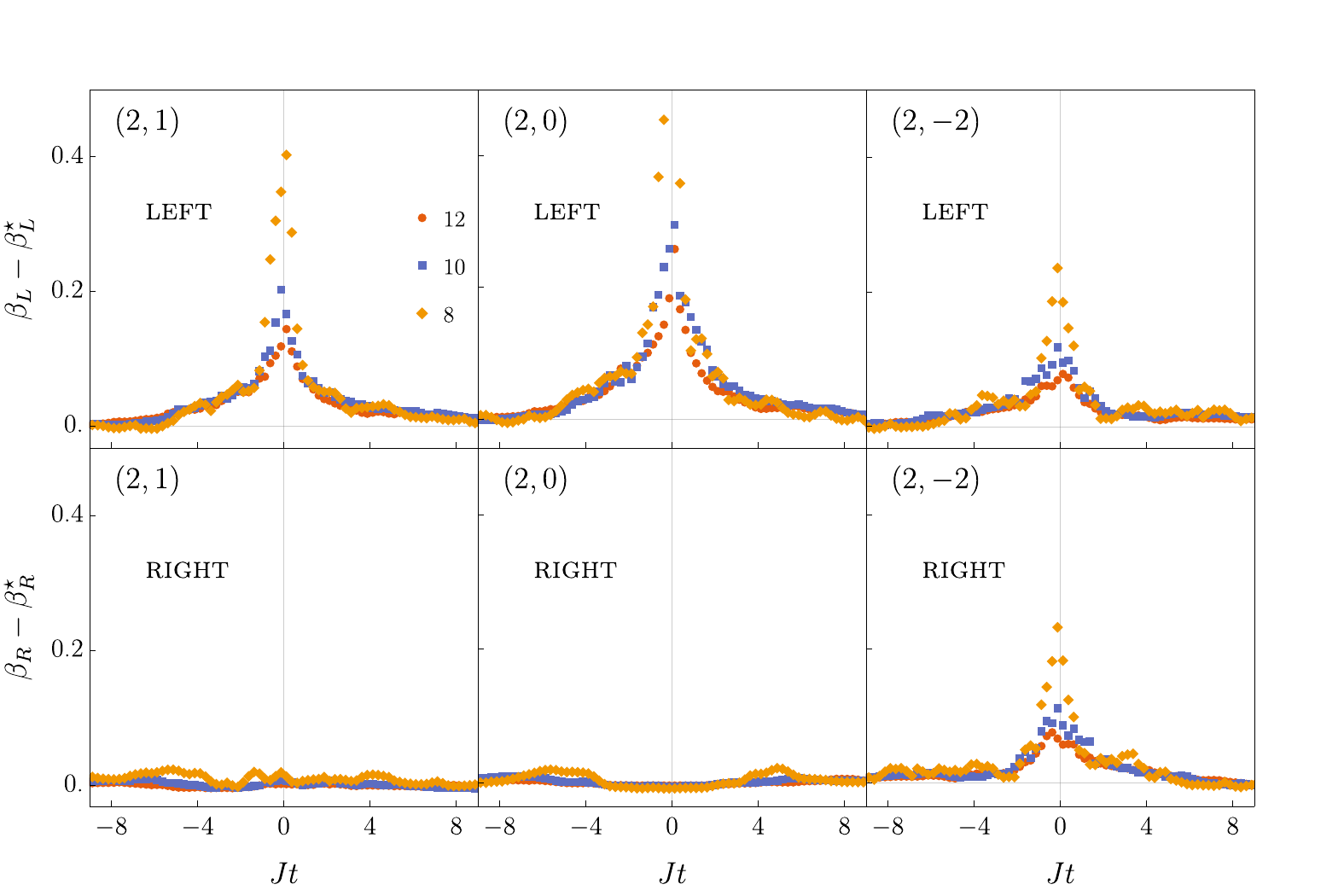}
\caption{\textbf{Bipartitioning protocol.} 
The half-chain inverse temperature after preparing the subchains at inverse temperature $(\beta_{0,L},\beta_{0,R})$ (shown on the upper-left corner of the plots in units of $J_0^{-1}$) for the Hamiltonian $H_{0,p}^{(I)}$ in \eqref{eq:H0pf}. The Hamiltonian is the same as in Figure~\ref{f:example1} with the additional interaction \eqref{eq:exampleV} with $\delta_y=0.3$. The symbols correspond to different chain lengths (in the legend). The half-chain inverse temperatures are shown relative to unspecified additive constants $\beta^\star_{L/R}$, since the assumptions required to define reliable anchors are not satisfied at these sizes.
}\label{f:example2}
\end{figure*}
 
We consider here the paradigm of transport: two systems associated with  disjoint sets of sites $I_{\textsc{l}}$ and $I_{\textsc{r}}$  ($I_{\textsc{l}}\cap I_{\textsc{r}}=\emptyset$), with  $-1\in I_{\textsc{l}}$, $0\in I_{\textsc{r}}$, and $|I_{\textsc{l}}|=|I_{\textsc{r}}|=L/2$ are prepared in equilibrium for given pre-quench Hamiltonians $H^{(I_{\textsc{l}})}_{0,\textsc{l}}$ and $H^{(I_{\textsc{r}})}_{0,\textsc{r}}$ with open boundary conditions.  The chains are joined, and the full system,  associated with the  set  of sites $I_{\textsc{l}}\cup I_{\textsc{r}}$, time evolves with the post-quench  Hamiltonian $H^{(I_{\textsc{l}}\cup I_{\textsc{r}})}$. We consider the simplified setting in which the two parts of the system are in equilibrium at different temperatures for the same pre-quench Hamiltonian $H_{0,p}^{(I_{\textsc{l}}/I_{\textsc{r}})}$ defined in \eqref{eq:H0pf}. We set the post-quench Hamiltonian to $H^{(I)}+V^{(I)}$, with the operators defined in \eqref{eq:exampleH} and \eqref{eq:exampleV} with open boundary conditions and choose generic coupling constants. 
In practice, we use exact diagonalization to construct the relevant operators at each time, including those of Eqs~\eqref{eq:HAHtildeA} and \eqref{eq:DH}. The time derivative in Eq.~\eqref{eq:firstlaw}  is then evaluated by discretization and interpolation, and the resulting rate integrated in time. For sufficiently large subsystems, the initial equilibrium preparation would provide a weak-canonical-flow anchor for the inverse temperatures of the two halves, with the intrinsic indetermination $O(1/L)$ discussed in Sections~\ref{ss:equi} and~\ref{ss:lowsub}. Whether alternative anchors exist at late times is less clear. At the accessible sizes, however, the assumptions required for reliable anchors cannot be established. We therefore refrain from estimating the absolute inverse temperatures and show only their variations, up to unspecified additive constants---Figure~\ref{f:example2}. The time range is extended to negative values to make visible how exceptional the state at $t=0$ is, a feature that hints at its ability to support a weak canonical flow. Figure~\ref{f:example1} shows that the larger $|\beta_0 J_0|$, the larger $\beta_{\omega_p}^H$ (note that, because of the spin-flip symmetry in the $y$ direction, $\beta_{\omega_p}^H$ does not depend on the sign of  $\beta_0 J_0$). The numerical data displayed in Figure~\ref{f:example2} therefore suggest that the temperature of the part of the chain with the largest temperature is approximately stationary.   In contrast, the temperature of the other part exhibits a clear tendency to increase. We note, however, that this temperature also appears to approach a plateau on relatively short timescales, with no clear indication that the corresponding timescale grows with $L$. The accessible data may therefore not probe the longer-time regime relevant to a possible approach of the two parts to a common temperature.  We also remark that, in the quench shown in the last column of Figure~\ref{f:example2}, the two parts have the same nonequilibrium temperature by symmetry. Interestingly, their temperature time evolves and the data are consistent with the two parts having roughly the same temperature at every time (note that the Hamiltonian is not reflection symmetric, so this cannot be explained by reflection symmetry). We refrain from drawing further conclusions and postpone a more detailed analysis to future work. 
\section{Discussion and outlook}

We have developed a thermodynamic framework that extends equilibrium thermodynamics to isolated quantum many-body systems out of equilibrium. 
We have adopted a geometric viewpoint in which the inverse temperature of a nonequilibrium isolated state is the coordinate of a canonical thermodynamic flow at fixed energy coherence. 
This parallels the structure of equilibrium thermodynamics without defining nonequilibrium temperature by comparison with equilibrium states. The parallel also extends to the formal level: the framework presented here naturally generalizes that of Ref.~\cite{Doyon2017Thermalization}, based on pseudolocal flows. At the same time, it requires a refinement of the usual maximum-entropy principle. The geometric viewpoint also clarifies when the generic one-temperature picture breaks down. 
In particular, we have shown that the same Hamiltonian can support both generic and non-generic leaves, whether or not it possesses additional integrals of motion. In non-generic cases the canonical temperature direction has to be supplemented by additional pseudolocal thermodynamic directions, leading to leaf grand-canonical ensembles or leaf generalized Gibbs ensembles, in analogy with the role of additional local and quasilocal charges in integrable systems~\cite{RigolDunjkoYurovskyOlshanii2007,EsslerFagotti2016, VidmarRigol2016,IlievskiDeNardisWoutersCauxEsslerProsen2015, IlievskiMedenjakProsenZadnik2016,Fagotti2017Charges}. Related mechanisms can also lead to fragmented thermodynamic sectors, as in systems with Hilbert-space fragmentation or nontrivial commutant algebras~\cite{SalaRakovszkyVerresenKnapPollmann2020,Zadnik2021The,Zadnik2021The2,MoudgalyaBernevigRegnault2022,MoudgalyaMotrunich2022}. We have deliberately overlooked topological structures on the leaves, which could manifest in the presence of pseudolocal flows associated with semilocal charges~\cite{Fagotti2022Global,FagottiMaricZadnik2024,Maric2023Semilocal}. Furthermore, we have touched on the issue of weak integrability breaking on the leaves only indirectly and very concisely, when discussing the fate of independent pseudolocal charges at later times. In view of the distinction between robust and fragile conserved quantities under perturbations~\cite{BurgarthFacchiNakazatoPascazioYuasa2021}, of recent constructions and obstructions for deforming integrable charges into quasi-conserved quantities under weak perturbations~\cite{BertiniEsslerGranet2022,KurlovMalikisGritsev2022,OrlovTiutiakinaSharipovPetrovaGritsevKurlov2023,SuraceMotrunich2023}, and of experimental signatures of fragile conservation laws in weakly broken integrable dynamics~\cite{ChenCapizziMarcheBornetEmperaugerLahayeBrowaeysFagottiMazza2026}, this appears to be a particularly natural direction for future investigation.

The subsystem construction also clarifies the distinction between temperature and equilibration. Temperature is a coordinate at fixed energy coherence, whereas equilibration is governed by the loss, redistribution, or local irrelevance of that coherence. A temperature-based distance therefore cannot, in general, parametrize the distance from equilibrium. From this perspective, Mpemba-like inversions are not exceptional violations of thermodynamic intuition; they instead reflect the fact that relaxation is being ordered by a parameter that is not aligned with the variables controlling the approach to equilibrium~\cite{KlichRazHirschbergVucelja2019,Moroder2024Thermodynamics,Vu2025Thermomajorization}.  We also showed that, although the instantaneous rate of change of the subsystem inverse temperature can become a local quantity, its value requires a reference point along the trajectory, depriving temperature of its intuitive status as an intrinsic property of the (subsystem) state. Moreover, the locality of the rate itself relies on an alignment assumption whose validity remains to be established.

Taken together, these questions point to a broader programme. A fully rigorous formulation of the framework should be developed directly in the $C^\star$-algebraic language of quasilocal observables, making the assumptions on regularity, clustering, and pseudolocality explicit. This would also make it possible to address the important question of extending the framework to higher-dimensional lattice systems, which exhibit a richer structure, including, for example, distinct phases of matter at nonzero temperature. Another aspect left for future investigation is the operational meaning of nonequilibrium temperature. In this respect, the integral representations exhibited for the $\omega$-dressed Hamiltonian and the harmonic conjugate of the Hamiltonian suggest a route to their numerical computation using state-of-the-art tensor-network algorithms, including matrix-product and tree-tensor-network methods~\cite{Banuls2023TensorNetworkAlgorithms}, especially when the state has Gibbs form with respect to the pre-quench Hamiltonian. A similar conclusion applies to the canonical flow, which connects regular states on a given leaf. Developments in this direction might provide an important step towards designing experimental protocols, particularly on quantum platforms, for measuring nonequilibrium temperature and the other thermodynamic quantities arising in the proposed framework. On the theoretical side, non-generic leaves require a systematic theory of their additional pseudolocal directions. In addition, the subsystem equation derived here suggests that temperature profiles may themselves become emergent dynamical fields on mesoscopic space-time scales, which, for integrable Hamiltonians, might exhibit interesting connections with generalized hydrodynamics~\cite{Essler2023Ashort}. Finally, by focusing on full-rank density matrices, we have left aside the possible existence of lower-dimensional orbits of regular states, which could provide a generalization of the concept of quantum many-body scars~\cite{MoudgalyaBernevigRegnault2022}. These directions share the same underlying message: away from equilibrium, quantum coherence is not an obstruction to thermodynamics, but part of the information that determines how thermodynamics is organized.

\begin{acknowledgments}
I thank Florent Ferro and Gianluca Morettini for discussions. I also thank Fabian Essler and Luca Tagliacozzo for useful comments.
\end{acknowledgments}
\appendix

\section{Coherent energy fluctuations and transport on the leaf}\label{ss:QFI}

We clarify here what remains, in the thermodynamic limit, of the connection between the leaves  and the QFI. Strictly speaking, the question cannot be phrased in the same way as in finite volume. First, the Hamiltonian of an infinite chain is not an observable of the quasilocal algebra. Second,  the finite-volume quantum Fisher information is extensive in regular states and therefore diverges in the thermodynamic limit. For this reason, the thermodynamic-limit object should not be the finite-volume QFI itself.  One possibility, often useful in translationally invariant many-body systems, is to consider the QFI per unit volume, in the spirit of Ref.~\cite{Hauke2016Measuring}. Here we take a more local point of view. We define directly a QFI density associated, as explained below, with the Hamiltonian density $h_\ell$. 

We start from the standard expression of the quantum Fisher information in terms of the symmetric logarithmic derivative~\cite{Liu2016Quantum}. For the unitary deformation generated by $H^{(I)}$, this gives
\begin{equation}\label{eq:QFIstandard}
F_Q(\rho_\omega;H^{(I)})=
\mathrm{tr}\left[
	\rho_\omega\left(L_{\rho_\omega}[\dot \rho_\omega]\right)^2
\right]=
4\,\langle\!\langle\tilde H_{[\omega]}^{(I)},\tilde H_{[\omega]}^{(I)}\rangle\!\rangle^c_\omega .
\end{equation}
We now express $\tilde H_{[\omega]}^{(I)}$ as $\sum_{\ell\in I}\tilde h_{(\omega'),\ell}^{(I)}$ where $\tilde h_{(\omega'),\ell}^{(I)}$ are the local densities~\eqref{eq:harmonicdensity}. Applying the leaf condition \eqref{eq:leaf-cond-3} with $O=2\tilde h_{(\omega'),\ell}$ gives
\begin{equation}
4\, \langle\!\langle\tilde H_{[\omega]}^{(I)},\tilde h^{(I)}_{(\omega'),\ell}\rangle\!\rangle^c_\omega=
2\,\omega\bigl(i[H^{(I)},\tilde h^{(I)}_{(\omega'),\ell}]\bigr)\, .
\end{equation}
It then follows that the finite-volume QFI admits the exact local decomposition
\begin{equation}
F_Q\Bigl(\rho_\omega;\sum_{\ell\in I} h_\ell\Bigr)=\sum_{\ell\in I}\mathcal F_Q^{(\omega')}(\omega;h_\ell)\, ,
\end{equation}
where the density on the right-hand side is given by
\begin{equation}\label{eq:QFIapp}
\mathcal F_Q^{(\omega')}(\omega;h_\ell)=2\,\omega\!\left(i[H,\tilde h_{(\omega'),\ell}]\right)\, ,
\end{equation}
and is well defined also in the thermodynamic limit $I\nearrow\mathbb Z$ (hence we dropped $I$ from the equation). Thus the local density is the primary object in the infinite-volume formulation. In translation-invariant situations, its spatial average gives the usual QFI per unit volume; in inhomogeneous situations, it remains a local density representative, in the same sense in which $h_\ell$ is a local representative of the Hamiltonian. Note that the dependence on $\omega'$ reflects a gauge freedom in the choice of the density representative. This is analogous to the familiar gauge freedom in the definition of local charge densities and currents: different representatives give the same extensive quantum Fisher information.

The analogy with conserved charges goes further. Since $F_Q(\rho_\omega;H^{(I)})$ is invariant under the unitary time evolution generated by $H^{(I)}$, its density satisfies a continuity equation. With the convention
\begin{equation}
i[h_\ell,H]=\hat J_{\ell}^{H}[h]-\hat J_{\ell-1}^{H}[h],
\end{equation}
and denoting by $\tilde J^{H}_{(\omega'),\ell}$ the harmonic conjugate of the energy current $\hat J_{\ell}^{H}[h]$, the associated QFI current is represented by
\begin{equation}\label{eq:QFIcurrent}
\mathcal J_Q^{(\omega')}(\omega;h_\ell)=2\,\omega\!\left(i[H,\tilde J^{H}_{(\omega'),\ell}]\right).
\end{equation}
Thus, for the time-evolving state $\omega_t=\omega\circ\sigma^t$, we obtain the continuity equation
\begin{equation}\label{eq:continuityQFI}
\partial_t\mathcal F_Q^{(\omega_t')}(\omega_t;h_\ell)=\mathcal J_Q^{(\omega_t')}(\omega_t;h_{\ell-1})-\mathcal J_Q^{(\omega_t')}(\omega_t;h_\ell)\, ,
\end{equation}
where $\omega_t'=\omega'\circ\sigma^t$ and $\omega'\sim_{\mathcal L_H}\omega $. In contrast to ordinary charge continuity equations, however, the operators entering the expectation values are state dependent and, along the physical time evolution, are transported with the leaf. Their typical range grows in time as the range of $\sigma^{-t}(h_\ell)$. If one restricts these operators to a fixed subsystem $A$, the part of the coherent-energy density carried outside $A$ is lost from the reduced description. From the subsystem viewpoint, this provides a local manifestation of the delocalization of coherent energy fluctuations, and hence of decoherence.

We shall not pursue the continuity equation~\eqref{eq:continuityQFI} and the associated QFI current~\eqref{eq:QFIcurrent} further in this work.  It would be interesting to investigate whether they can be used to characterize inhomogeneous states, where the local redistribution of coherent energy fluctuations may carry information not captured by spatially averaged quantities (see also Section~\ref{s:examples}).

\section{Minimum Discrimination Information versus Maximum Entropy}\label{a:principle}
We clarify here the origin of the tension between the principle of
minimum discrimination information and the principle of maximum entropy.
To focus on the essential point, we consider a paradigmatic leaf with a
single pseudolocal charge, associated with the Hamiltonian, and supporting strong leaf
typicality, i.e., all pure states of the minimum-variance optimal
decomposition with the same energy density as $I\nearrow\mathbb Z$ are
locally equivalent in the thermodynamic limit.

Strong leaf typicality implies that the leaf canonical ensemble
introduced in Ref.~\cite{Fagotti2026Quantum},
\begin{equation}\label{app:leafGibbs}
\omega_{\bullet,\beta}=\frac{
\rho_{\omega_\bullet}^{\frac12}
e^{-\beta H_{(\omega_\bullet)}^{(I)}}
\rho_{\omega_\bullet}^{\frac12}
}{
\mathrm{tr}\!\left[
\rho_{\omega_\bullet}
e^{-\beta H_{(\omega_\bullet)}^{(I)}}
\right]
},
\end{equation}
is locally equivalent, at the same energy density, to the leaf canonical
ensemble through $\omega$, Eq.~\eqref{eq:rhoomegadef},
\begin{equation}
\omega_{\beta_\omega^H+\Delta\beta}=\frac{
\rho_\omega^{\frac12}
e^{-\Delta\beta H_{(\omega)}^{(I)}}
\rho_\omega^{\frac12}
}{
\mathrm{tr}\!\left[
\rho_\omega
e^{-\Delta\beta H_{(\omega)}^{(I)}}
\right]
}.
\end{equation}
The local equivalence, however, is in general \emph{not} realized by the
trivial identification
\begin{equation}\label{app:trivialshift}
\beta=\beta_\omega^H+\Delta\beta .
\end{equation}
Remarkably, this can happen even though Eq.~\eqref{eq:condT0}
effectively enforces precisely such an identification at the
infinite-temperature macrostate, as made explicit in
Section~\ref{ss:nongeneric}.

Both constructions use the same canonical response: under a variation of
inverse temperature along the canonical flow, the variation of the energy
density is fixed, up to a minus sign, by the incoherent part of the energy
fluctuations, cf.~\eqref{eq:betainner}. In the maximum-entropy construction in which the states on the leaf are qualified only geometrically as points on the simplex, the canonical flow starts from the
barycenter $\omega_\bullet$, which has uniform population. At the value of $\beta$ for which the energy
density coincides with that of $\omega$, the flow reaches a state that,
by strong leaf typicality, is locally equivalent to $\omega$. In general,
however, this state does not coincide with $\omega$.
Minimum discrimination information reverses the construction. The
canonical flow is now anchored at the specific state $\omega$ whose
temperature is to be determined. Following the opposite flow by
$-\beta_\omega^H$ reaches a state with the energy density of the
barycenter and hence, under strong leaf typicality, locally equivalent to
$\omega_\bullet$. 
If we call $\beta_{\omega'}^{m\textsc{di}}$ the inverse temperature of $\omega'\in\mathcal L_H(\omega)$ obtained with the principle of minimum discrimination information ($\beta_{\omega'}^{m\textsc{di}}\equiv \beta_{\omega'}^{H}$) and $\beta_{\omega'}^{ \textsc{me}}$ the inverse temperature obtained with the principle of maximum entropy, matching the energy difference of $\omega'$ with that of the barycenter $\omega_\bullet$ gives 
\begin{equation}
\frac{d \beta_{\omega'}^{\textsc{me}}}{d \beta_{\omega'}^{m\textsc{di}}}  =\lim_{I\nearrow\mathbb Z} \frac{\|H^{(I)}\|_{\omega',{\rm inc}}^2}{\|H^{(I)}\|_{\omega_{\bullet,\beta_{\omega'}^{\textsc{me}}},{\rm inc}}^2}\, .
\end{equation}
Crucially, the right-hand side is not necessarily equal to $1$: while, by leaf typicality, $\omega'$ and $\omega_{\bullet,\beta_{\omega'}^{\textsc{me}}}$ have the same density of QFI with respect to $H$, they can have different energy-variance densities and hence, by \eqref{eq:incoherentnorm}, different densities of the squared incoherent norm of the Hamiltonian. 

The apparent tension between maximum entropy and minimum discrimination information is therefore a consequence of their different anchoring. That choice, however, is not merely conventional: the pseudolocality of the canonical flow depends on the properties of its anchor. When the state is given, minimum discrimination information is therefore the more natural principle. Note that the principle of minimum discrimination information can be rephrased as a principle of maximum entropy if we agree to choose the population of $\omega$ (or of any other state in the orbit of the canonical flow through it) as a prior. We avoided this terminology to highlight the importance of the state and to circumvent the inverse problem of finding a regular thermodynamic state for a leaf specified by one of the leaf conditions presented in Section~\ref{s:leaves}, e.g., by the density of $H_{[\omega]}^{-}$.

\section{On the regularity of the canonical flow}\label{a:leafthrough}

In this appendix we provide a cluster-expansion argument showing that, for sufficiently small values of the nonequilibrium inverse-temperature coordinate, the leaf-canonical flow through a regular KMS state is analytic on local observables and generates exponentially clustering states. Specifically, we adapt the high-temperature cluster-expansion results of Ref.~\cite{NguyenFernandez2024HighTemperature} to the setting considered here. For concreteness, we restrict ourselves to spin-$\frac{1}{2}$ chains.

\subsection{Cluster expansion}

We first summarize the definitions and results of Ref.~\cite{NguyenFernandez2024HighTemperature} used below. The decoupling-parameter construction underlying that expansion goes back to Park~\cite{Park1982ClusterExpansion}. Given a finite-volume interaction
\begin{equation}
H=\sum_X\Phi(X),
\end{equation}
each finite set $X$ such that $\Phi(X)\neq0$ is called a \emph{bond}. A finite family of bonds $\mathcal B$ has physical support
\begin{equation}
\underline{\mathcal B}=\bigcup_{X\in\mathcal B}X.
\end{equation}
Its overlap graph has one vertex for each bond and an edge between bonds with intersecting supports. 
A nonempty bond family whose overlap graph is connected is called a \emph{polymer}; two polymers are compatible when their physical supports are
disjoint and incompatible otherwise. Given a finite sequence of polymers $(P_1,\ldots,P_n)$, its incompatibility graph has vertex set $\{1,\ldots,n\}$ and an edge between $i$ and $j$ whenever $P_i$ and $P_j$ are incompatible. A \emph{cluster} is a finite sequence of polymers whose incompatibility graph is connected.

For a bond family $\mathcal B$, the restricted Boltzmann weight and its operator-valued M\"obius coefficient are
\begin{equation}\label{eq:appendix-reference-Mobius}
U(\mathcal B)=\exp\!\left[-\sum_{X\in\mathcal B}\Phi(X)\right],\quad\xi(\mathcal B)=
\sum_{\mathcal C\subseteq\mathcal B}(-1)^{|\mathcal B\setminus\mathcal C|}U(\mathcal C).
\end{equation}
The coefficient $\xi(\mathcal B)$ isolates the contribution that depends jointly on all bonds in $\mathcal B$. Since operators with disjoint supports commute, $U(\mathcal B)$ factorizes over compatible bond families. M\"obius inversion then implies that $\xi(\mathcal B)$ factorizes over the connected components of $\mathcal B$. Proposition~6 of Ref.~\cite{NguyenFernandez2024HighTemperature} turns any such factorizing weight into an operator-valued hard-core gas of polymers; applied to $U(\mathcal B)$, this is Proposition~9 of the same reference, which expands the full Boltzmann weight.

Taking the normalized trace
$\tau_{\underline P}=2^{-|\underline P|}
\mathrm{tr}_{\underline P}$ over the physical support of the polymer
gives a scalar polymer gas with activities
\begin{equation}
w(P)=\tau_{\underline P}\bigl(\xi(P)\bigr).
\end{equation}
The logarithm of the partition function is expanded in clusters, while pinned expansions restrict the sum to clusters containing a prescribed polymer or meeting a prescribed finite support.

The results of the reference that we need are the following. Lemma~24 gives the bondwise estimate
\begin{equation}\label{eq:appendix-reference-majorant}
\left\|\xi(\mathcal B)\right\|\leq\prod_{X\in\mathcal B}\left(\mathrm e^{\|\Phi(X)\|}-1\right).
\end{equation}
In the high-temperature region considered there, the cluster and pinned-cluster expansions converge absolutely and uniformly in the volume.  Moreover, Proposition~11 and Corollary~36 express normalized expectations of local observables as sums of connected families attached to their supports and allow the thermodynamic limit to be taken term by term. For a general treatment of cluster expansions and correlation functions, see also Ref.~\cite{Ueltschi2004ClusterExpansions}.

We assume that the interaction defining the state $\omega$ lies strictly inside this convergence region. In particular, the same bounds are assumed to remain valid under sufficiently small changes of the activities and after multiplying each one-bond majorant by $\mathrm e^{\mu|X|}$ for some $\mu>0$. We now verify that the construction carries over to the ordered product of two exponentials appearing in the leaf-canonical flow.

\subsection{Layer-labelled polymer expansion}

Let $I\subset\mathbb Z$ be a finite interval and set
\begin{equation}\label{eq:appendix-reference-state}
\rho_\omega^{(I)}=\frac{\mathrm e^{-\beta_0H_0^{(I)}}}{\mathrm{tr}_I\!\left(\mathrm e^{-\beta_0H_0^{(I)}}\right)}\, .
\end{equation}
By construction, the operators $H_{[\omega]}^{(I)\pm}$ introduced in the main text satisfy the finite-volume leaf condition \eqref{eq:leaf-intertwine}. Iterating the intertwining relation and summing the exponential series therefore gives
\begin{equation}\label{eq:appendix-finite-leaf-relation}
\mathrm e^{-sH_{[\omega]}^{(I)-}}\rho_\omega^{(I)}=\rho_\omega^{(I)}\mathrm e^{-sH_{[\omega]}^{(I)+}},
\qquad s\in\mathbb C\, .
\end{equation}
Consequently, for every local observable $O$ supported in $I$, cyclicity of the trace yields
\begin{multline}\label{eq:appendix-F-finite}
F_O^{(I)}(z)=
\tfrac{
	\mathrm{tr}_I\!\bigl[\rho_\omega^{(I)}\mathrm e^{-zH_{[\omega]}^{(I)+}/2}O\mathrm e^{-zH_{[\omega]}^{(I)-}/2}\bigr]
}{
	\mathrm{tr}_I\!\bigl[\rho_\omega^{(I)}\mathrm e^{-zH_{[\omega]}^{(I)+}/2}\mathrm e^{-zH_{[\omega]}^{(I)-}/2}\bigr]
}\\
=
\tfrac{
	\tau_I\!\bigl(O\mathrm e^{-zH_{[\omega]}^{(I)-}}\mathrm e^{-\beta_0H_0^{(I)}}\bigr)
}{
	\tau_I\!\bigl(\mathrm e^{-zH_{[\omega]}^{(I)-}}\mathrm e^{-\beta_0H_0^{(I)}}\bigr)
}\, .
\end{multline}
For real $z$, using $H_{[\omega]}^{(I)+}=(H_{[\omega]}^{(I)-})^\dagger$, the first expression in \eqref{eq:appendix-F-finite} is the expectation value of $O$ in the state with density matrix
\begin{equation}\label{eq:appendix-rho-z}
\rho_{z,I}=
\frac{
	\mathrm e^{-zH_{[\omega]}^{(I)-}/2}\rho_\omega^{(I)}\mathrm e^{-zH_{[\omega]}^{(I)+}/2}
}{
	\mathrm{tr}_I\!\left[\mathrm e^{-zH_{[\omega]}^{(I)-}/2}\rho_\omega^{(I)}\mathrm e^{-zH_{[\omega]}^{(I)+}/2}\right]
}\, .
\end{equation}
We define the interactions $\Phi_0^{(I)}$ and $\Phi_1^{(I)}$ such that
\begin{equation}\label{eq:appendix-interactions}
\beta_0H_0^{(I)}=\sum_{\emptyset\neq X\subset I}\Phi_0^{(I)}(X),
\quad H_{[\omega]}^{(I)-}=\sum_{\emptyset\neq X\subset I}\Phi_1^{(I)}(X)\, ,
\end{equation}
where the nonzero terms have interval supports. We assume that the two families, which we call \emph{layer-$0$} and \emph{layer-$1$}, are uniformly exponentially summable\footnote{For the layer-$1$ family, this assumption follows from the integral representation of $h_{(\omega),\ell}^{\pm}$ whenever the reference dynamics satisfies a uniform Lieb--Robinson bound. Indeed, writing $\mathcal L_0(A)=[H_0,A]$ and using the Fourier transform of $1/\cosh(\pi s)$ gives
\[
h_{(\omega),\ell}^{\pm}=\left[1\pm\tanh\!\left(\frac{\beta_0\mathcal L_0}{2}\right)\right]h_\ell .
\]
Equivalently,
\begin{multline*}
\widetilde h_{(\omega),\ell}:=\frac{i}{2}\left(h_{(\omega),\ell}^{-}-h_{(\omega),\ell}^{+}\right)=\\
\frac{i}{\pi}\int_{-\infty}^{\infty}dt\,\log\coth\!\left(\frac{\pi|t|}{2\beta_0}\right)\sigma^t_0\!\left([h_\ell,H_0]\right).
\end{multline*}
The kernel is integrable at the origin and decays exponentially for large $|t|$. Combining this decay with the Lieb–Robinson approximation of a time-evolved local observable by one supported in its effective light cone~\cite{BravyiHastingsVerstraete2006} yields exponential quasilocality of $h_{(\omega),\ell}^{\pm}$, with a rate depending on $\beta_0$ and on the Lieb–Robinson velocity. A telescopic decomposition into successive interval-supported local approximants then yields an exponentially summable interaction $\Phi_1^{(I)}$.} and that, for every fixed interval $X$,
\begin{equation}\label{eq:appendix-local-convergence}
\Phi_a^{(I)}(X)\longrightarrow\Phi_a(X)
\quad\text{as }I\nearrow\mathbb Z
\qquad (a=0,1)
\end{equation}
in operator norm. 
We also assume that the finite-volume reference states \eqref{eq:appendix-reference-state} converge locally to $\omega$.

To preserve the order of the two exponentials, we introduce \emph{layer-labelled bonds}
\begin{equation}
\mathfrak b=(a,X), 
\qquad a\in\{0,1\},
\qquad \underline{\mathfrak b}=X.
\end{equation}
Two labelled bonds are incompatible when their physical supports intersect. For a finite family $\mathcal B$ of labelled bonds, let
\begin{equation}
\mathcal B_a=\left\{(a,X)\in\mathcal B\right\},
\qquad\underline{\mathcal B}=\bigcup_{(a,X)\in\mathcal B}X,
\end{equation}
and define the ordered partial weight
\begin{multline}\label{eq:appendix-ordered-partial-weight}
U_{z,I}(\mathcal B)=\\
\exp\Bigl[-z\sum_{(1,X)\in\mathcal B_1}\Phi_1^{(I)}(X)\Bigr]\exp\Bigl[-\sum_{(0,X)\in\mathcal B_0}\Phi_0^{(I)}(X)\Bigr].
\end{multline}
If $\underline{\mathcal B}\cap\underline{\mathcal C}=\emptyset$, every operator belonging to $\mathcal B$ commutes with every operator belonging to $\mathcal C$, and therefore
\begin{equation}\label{eq:appendix-disjoint-factorization}
U_{z,I}(\mathcal B\cup\mathcal C)=U_{z,I}(\mathcal B)U_{z,I}(\mathcal C)
\qquad \underline{\mathcal B}\cap\underline{\mathcal C}=\emptyset.
\end{equation}
Thus the factorization property required by the operator-gas construction is unchanged.

The corresponding M\"obius coefficient is
\begin{equation}\label{eq:appendix-Mobius-coefficient}
\xi_{z,I}(\mathcal B)=\sum_{\mathcal C\subseteq\mathcal B}(-1)^{|\mathcal B\setminus\mathcal C|}U_{z,I}(\mathcal C).
\end{equation}
Since the subsets of the two layers can be summed independently,
\begin{equation}\label{eq:appendix-layer-factorization}
\xi_{z,I}(\mathcal B)=\xi_{z,I}^{(1)}(\mathcal B_1)\xi_I^{(0)}(\mathcal B_0),
\end{equation}
where the two factors are the corresponding one-layer M\"obius coefficients. 

The estimate of Lemma~24 of
Ref.~\cite{NguyenFernandez2024HighTemperature} is obtained by repeatedly
using
\begin{equation}
\mathrm e^{A+B}-\mathrm e^A
=
\int_0^1 d\lambda\,
\mathrm e^{(1-\lambda)(A+B)}B\mathrm e^{\lambda A},
\end{equation}
together with the operator-norm inequality
$\|AB\|\leq\|A\|\,\|B\|$. These relations hold for arbitrary bounded
operators, so the proof does not require Hermiticity and remains valid for
the layer-$1$ terms and complex $z$. Applying the estimate separately to
the two factors in \eqref{eq:appendix-layer-factorization} gives
\begin{multline}\label{eq:appendix-Mobius-bound}
\bigl\|\xi_{z,I}(\mathcal B)\bigr\|\leq\\
\prod_{(1,X)\in\mathcal B_1}\bigl(\mathrm e^{|z|\|\Phi_1^{(I)}(X)\|}-1\bigr)
\prod_{(0,X)\in\mathcal B_0}\bigl(\mathrm e^{\|\Phi_0^{(I)}(X)\|}-1\bigr).
\end{multline}

A polymer is now a nonempty connected family of labelled bonds. By Proposition 6 of Ref.~\cite{NguyenFernandez2024HighTemperature}, \eqref{eq:appendix-disjoint-factorization} gives the corresponding operator-valued hard-core gas expansion. Applying the normalized product trace gives the scalar activities
\begin{equation}\label{eq:appendix-polymer-fugacity}
w_{z,I}(\mathcal P)=\tau_{\underline{\mathcal P}}\!\left(\xi_{z,I}(\mathcal P)\right),
\qquad \left|w_{z,I}(\mathcal P)\right|\leq\left\|\xi_{z,I}(\mathcal P)\right\|.
\end{equation}

The bound \eqref{eq:appendix-Mobius-bound} has the form required by the convergence criteria of Ref.~\cite{NguyenFernandez2024HighTemperature}, with each pair $(a,X)$ regarded as a distinct bond. Since the layer-$1$ majorants vanish at $z=0$ and depend continuously on $z$, the assumed strict margin implies that there exists $\Delta\beta_*>0$ such that the cluster and observable expansions converge absolutely and uniformly in $I$ on every closed disc $|z|\leq z_0<\Delta\beta_*$. In particular, the cluster expansion of the logarithm of the last denominator in \eqref{eq:appendix-F-finite} converges there, so the denominator does not vanish and $F_O^{(I)}(z)$ is analytic on the disc.

\subsection{Thermodynamic limit and clustering}

The proofs of Proposition~11 and Corollary~36 of Ref.~\cite{NguyenFernandez2024HighTemperature} carry over to the layer-labelled gas. The resulting expansion of \eqref{eq:appendix-F-finite} contains only connected families attached to $\mathrm{supp}(O)$. Its locally uniform convergence and \eqref{eq:appendix-local-convergence} imply that
\begin{equation}\label{eq:appendix-F-limit}
F_O(z)=\lim_{I\nearrow\mathbb Z}F_O^{(I)}(z)
\end{equation}
exists locally uniformly for $|z|<\Delta\beta_*$ and is analytic there. Since the finite-volume Gibbs representations of $\omega$ converge locally to $\omega$,
\begin{equation}
F_O(0)=\omega(O).
\end{equation}
For real $z$, Eq.~\eqref{eq:appendix-rho-z} shows that the finite-volume functionals are positive and normalized. These properties pass to the local thermodynamic limit, and hence
\begin{equation}\label{eq:appendix-relative-state}
\omega_z:O\mapsto F_O(z),
\qquad z\in\mathbb R,
\quad |z|<\Delta\beta_*,
\end{equation}
defines a state.

The same expansion also gives exponential clustering. Let $O_1$ and $O_2$ be local observables with disjoint supports and introduce
\begin{equation}\label{eq:appendix-source-partition-function}
Z_I(z;s,t) =\tau_I\!\left(\mathrm e^{sO_1}\mathrm e^{tO_2}\mathrm e^{-zH_{[\omega]}^{(I)-}}\mathrm e^{-\beta_0H_0^{(I)}}\right).
\end{equation}
Since $O_1$ and $O_2$ have disjoint supports, they commute, and
\begin{equation}\label{eq:appendix-connected-source}
\left.\partial_s\partial_t\log Z_I(z;s,t)\right|_{s=t=0}=
F_{O_1O_2}^{(I)}(z)-F_{O_1}^{(I)}(z)F_{O_2}^{(I)}(z).
\end{equation}
The preceding layer-labelled construction extends verbatim by treating $\mathrm e^{sO_1}$ and $\mathrm e^{tO_2}$ as two additional one-bond layers, supported on $\mathrm{supp}(O_1)$ and $\mathrm{supp}(O_2)$, respectively. In the expansion of $\log Z_I(z;s,t)$, the mixed derivative at $s=t=0$ selects precisely the clusters containing both additional bonds. Every such cluster therefore contains a connected chain of overlapping interval supports joining $\mathrm{supp}(O_1)$ to $\mathrm{supp}(O_2)$. By the assumed exponentially weighted convergence margin, the same cluster bounds remain valid after multiplying every interaction-bond majorant by $\mathrm e^{\mu|X|}$ for some $\mu>0$. Any connected chain $\Gamma$ of interaction bonds joining $\mathrm{supp}(O_1)$ to $\mathrm{supp}(O_2)$ satisfies
\begin{equation}
\sum_{(a,X)\in\Gamma}|X| \geq \mathrm{dist}(\mathrm{supp}(O_1),\mathrm{supp}(O_2))\, .
\end{equation}
The weighted cluster bound therefore gives, for every $z_0<\Delta\beta_*$,
\begin{multline}\label{eq:appendix-clustering}
\left|\omega_z(O_1O_2)-\omega_z(O_1)\omega_z(O_2)\right|\leq\\
C_{O_1,O_2,z_0}\mathrm e^{-\mu_{z_0}\mathrm{dist}\left(\mathrm{supp}(O_1),\mathrm{supp}(O_2)\right)}
\end{multline}
for real $|z|\leq z_0$, where $\mu_{z_0}>0$ and the bound is uniform under translations of the two observables. We also note that, by Ref.~\cite{Ampelogiannis2025Clustering}, the two-point estimate \eqref{eq:appendix-clustering} implies analogous exponential decay for every fixed-order connected correlation when one local observable is separated from the others.

Finally, this regularity regime is compatible with the determination of the absolute inverse temperature. As $\beta_0\rightarrow 0$, the state $\omega$ approaches the tracial state whereas $\Delta\beta_\ast$ remains bounded below by a positive constant. Since, by \eqref{eq:partomega}, the energy density is strictly decreasing on the flow, \eqref{eq:condT0} has a unique solution, which tends to zero in that limit; hence there is a range of $\beta_0$ in which $-\beta_\omega^H$  lies in the disc $|z|<\Delta\beta_\ast$. In that case, the canonical orbit joining 
$\omega$ to the reference state is regular throughout, and $\beta_\omega^H$ is well defined by \eqref{eq:condT0}.
Thus, throughout a sufficiently small high-temperature interval, the leaf-canonical flow is analytic on local observables and preserves exponential clustering under the stated assumptions.

\section{On the canonical flow induced on subsystems}\label{a:subsystem}

In this appendix, we justify Equation~\eqref{eq:betadot}. For the sake of simplicity, the notations will hide the time dependence, but the states and all the quantities derived from them should be understood as time dependent. 
The starting point is the canonical pseudolocal charge \eqref{eq:canonical}. Its restriction to observables supported in $A$ defines the linear functional
\begin{equation}\label{eq:restrictedfunctional}
O\mapsto \widehat H_\omega(O),
\qquad O\in\mathcal A_A,
\end{equation}
where $\mathcal A_A$ denotes the algebra of observables supported in $A$. We interpret this functional as the canonical direction of an effective subsystem problem. We therefore look for an effective subsystem Hamiltonian $H_A^\omega$ such that the canonical pseudolocal charge constructed from the pair $(\omega_A,H_A^\omega)$ coincides with the restriction of the global one:
\begin{multline}\label{eq:reduction}
\langle\!\langle H_A^\omega,O\rangle\!\rangle_{\omega_A}^{c}+\frac{i}{2}\omega_A\!\left([\tilde H^\omega_{A,[\omega_A]},O]\right)\equiv\\
\lim_{I\nearrow\mathbb Z}\left[\langle\!\langle H^{(I)},O\rangle\!\rangle_{\omega}^{c}+\frac{i}{2}\omega\!\left([\tilde H^{(I)}_{[\omega]},O]\right)\right]\, ,
\end{multline}
where $O\in\mathcal A_A$.
Here $\tilde H^\omega_{A,[\omega_A]}$ is the harmonic conjugate of $H_A^\omega$ with respect to the reduced state $\omega_A$. The left-hand side is therefore written entirely in terms of the effective subsystem data. 
The inversion of \eqref{eq:reduction} is most transparent using the general identity
\begin{equation}
\widehat K_\eta(O)=\langle K,O^-_{(\eta)}\rangle_{\eta}^{c},
\end{equation}
where $\eta$ is a state, $K$ is the operator generating the canonical functional, and $O^-_{(\eta)}=\lim_{\lambda\to-1/2}O_{(\eta)}(\lambda)$ is defined by the same analytic continuation as in \eqref{eq:hw-def}. By the Gibbs condition satisfied by $\omega$---see Section~\ref{ss:initialstate}---the reduced density matrix $\rho_{\omega_A}$ of every finite subsystem $A$ is invertible. Hence the map $O\mapsto O^-_{(\omega_A)}$ is a well-defined invertible linear map on the finite-dimensional algebra $\mathcal A_A$. The matching condition \eqref{eq:reduction} can therefore be rewritten as
\begin{equation}\label{eq:inversionreduction}
\langle H_A^\omega,O^{\omega,-}_{A,(\omega_A)}\rangle_{\omega_A}^{c}=\widehat H_\omega(O),
\qquad O\in\mathcal A_A.
\end{equation}
This determines $H_A^\omega$ up to an additive scalar, which may be fixed, for instance, by imposing $\tau(H_A^\omega)=0$, with $\tau$ the tracial state.  

The effective leaf obtained in this way should be understood only locally in state space, in a neighborhood of the reduced state, as depicted in Fig.~\ref{f:submap} bottom. Indeed, $H_A^\omega$ should not be regarded as a state-independent subsystem Hamiltonian. Even with the weak notion of canonical flow introduced in Section~\ref{ss:equi}---where only the action of the flow on observables supported in the bulk of $A$ is retained---the induced pseudolocal flow can depend on the regular full-system state $\omega_\beta$ lying on the canonical trajectory through $\omega$. The construction therefore does not define a global foliation of the subsystem state space, but rather an effective thermodynamic leaf adapted to the subsystem state under consideration.

For simplicity, we assume that $A$ is finite and that the corresponding dressed subsystem Hamiltonian $H^\omega_{A,(\omega_A)}$ has simple spectrum. The construction is then smooth along the physical trajectory, away from degeneracies. Indeed, the global state $\omega$ and its restriction $\omega_{A}$ have smooth expectation values on local observables. The restricted canonical functional \eqref{eq:restrictedfunctional} is smooth under the regularity assumptions entering the pseudolocal construction. Moreover, since time evolution is an automorphism, the finite-volume reduced states $\omega_{A}$ remain faithful. Hence $O\mapsto O^{\omega,-}_{A,(\omega_{A})}$ is a smooth family of invertible linear maps on $\mathcal A_A$. The inversion~\eqref{eq:inversionreduction} therefore determines $H_A^{\omega}$, up to an additive scalar, as a smooth function of time. If $H^{\omega}_{A,(\omega_{A})}$ remains nondegenerate, its spectral projectors, and hence the optimal decomposition defining the effective leaf, also vary smoothly with time. 

The remaining step is to compare effective leaves at neighbouring times. In finite volume, we parallel-transport the optimal decomposition while keeping the population coordinates fixed. This preserves the thermodynamic entropy and gives \eqref{eq:virtual}. As discussed in Section~\ref{ss:firstlaw}, the separation between the transported state $\omega_A^{\mathrm{vir}}$ and $\omega_A$ generally has components transverse to the canonical direction. Under the alignment assumption, their contribution to the variation of the thermodynamic entropy is subleading for a sufficiently large subsystem. We therefore identify $\omega_A^{\mathrm{vir}}$ with the corresponding point on the regular canonical orbit through $\omega_A$. The separation between the actual and virtual states is then an infinitesimal displacement $d\beta_A$ along that orbit, and
\begin{multline}
S_{\rm th}(\omega_{A}|\mathcal L_{H_A^\omega}(\omega_A))-S_{\rm th}(\omega^-_{A}|\mathcal L_{H_A^{\omega^-}}(\omega^-_A))=\\
S_{\rm th}(\omega_{A}|\mathcal L_{H_A^\omega}(\omega_A))-S_{\rm th}(\omega_{A}^{\rm vir}|\mathcal L_{H_A^{\omega}}(\omega_A))=\\
\partial_{\beta_A} S_{\rm th}(\omega_A|\mathcal L_{H_A^\omega}(\omega_A)) d\beta_A+o(dt),
\end{multline}
where $\omega^-$ is the state at time $t-dt$ and the first equality follows from \eqref{eq:virtual}. If the subsystem is large enough for the inverse temperature to admit a deterministic interpretation as a leaf coordinate, dividing by $dt$ and taking $dt\to0$ gives \eqref{eq:betadot}.

\bibliography{references.bib}

\end{document}